\documentclass[12pt,a4paper]{article}
\usepackage[dvips]{graphicx,color}
\newcommand{\be}{\begin{equation}}
\newcommand{\ee}{\end{equation}}
\newcommand{\ba}{\begin{eqnarray}}
\newcommand{\ea}{\end{eqnarray}}
\newcommand{\baa}{\begin{eqnarray*}}
\newcommand{\eaa}{\end{eqnarray*}}
\newcommand{\lab}[1]{\label{#1}}

\newcommand{\dis}{\displaystyle}
\newcommand{\biq}{\mbox{\boldmath $q$}}
\newcommand{\bip}{\mbox{\boldmath $p$}}

\newcommand{\biL}{\mbox{\boldmath $\Lambda$}}
\newcommand{\bil}{\mbox{\boldmath $\lambda$}}
\newcommand{\bisg}{\mbox{\boldmath $\sigma$}}

\newcommand{\bs}[1]{\mbox{\boldmath $#1$}}

\newcommand{\cP}{\mathcal{P}}
\newcommand{\cV}{\mathcal{V}}
\newcommand{\cM}{\mathcal{M}}
\newcommand{\cL}{\mathcal{L}}
\begin{document}
{\pagestyle{empty}
\vskip 1.5cm

{\renewcommand{\thefootnote}{\fnsymbol{footnote}}
\centerline{\large \bf ENHANCED SAMPLING ALGORITHMS}
}
\vskip 3.0cm
 
\centerline{Ayori Mitsutake,$^{a}$
Yoshiharu Mori,$^{b}$
and Yuko Okamoto$^{b,c,}$\footnote{\ \ Corresponding author.
e-mail: okamoto@phys.nagoya-u.ac.jp}}
\vskip 1.5cm
\centerline{$^a${\it Department of Physics}}
\centerline{{\it Keio University}}
\centerline{{\it Yokohama, Kanagawa 223-8522, Japan}}
%\centerline{and}
\centerline{$^b${\it Department of Physics}}
\centerline{{\it Nagoya University}}
\centerline{{\it Nagoya, Aichi 464-8602, Japan}}
%\centerline{and}
\centerline{$^c${\it Structural Biology Research Center}}
\centerline{{\it Nagoya University}}
\centerline{{\it Nagoya, Aichi 464-8602, Japan}}

\vskip 1.5cm

\centerline{{\it {\bf Keywords:} Monte Carlo; molecular dynamics; 
generalized-ensemble algorithm; replica-exchange method;}}
\centerline{{\it simulated tempering; multicanonical algorithm}}

\medbreak
\vskip 1.0cm
 
\centerline{\bf ABSTRACT}
\vskip 0.3cm

In biomolecular systems (especially all-atom models) 
with many degrees of freedom such as
proteins and nucleic acids, there exist an astronomically large
number of local-minimum-energy states.  Conventional simulations 
in the canonical
ensemble are of little use, because they tend to get trapped in
states of these energy local minima.
Enhanced conformational sampling techniques are thus in great
demand.
A simulation in generalized ensemble performs a random walk
in potential energy space and can overcome this difficulty.
From only one simulation run, one can
obtain canonical-ensemble averages of physical quantities as
functions of temperature
by the single-histogram and/or multiple-histogram reweighting techniques.
In this article we review uses of the generalized-ensemble algorithms in biomolecular systems.
Three well-known methods, namely, 
multicanonical algorithm, simulated tempering, and 
replica-exchange method, are described first.
Both Monte Carlo and molecular dynamics versions of the 
algorithms are given.
We then present various extensions of these
three generalized-ensemble algorithms.
The effectiveness of 
the methods 
%for molecular simulations in the protein folding problem 
is tested with short peptide and protein systems.

%\vfill
%\newpage}
}
%\baselineskip=0.8cm
%%%%%%%%%%%%%%%%%%%%%%% Section 1 %%%%%%%%%%%%%%%%%%%%%%%%
%\noindent
%{\bf 1. INTRODUCTION} \\
\section{INTRODUCTION}

Conventional Monte Carlo\index{Monte Carlo} (MC)\index{MC} and
molecular dynamics\index{molecular dynamics} (MD)\index{MD} 
simulations of biomolecules 
are greatly hampered by the multiple-minima
problem\index{multiple-minima problem}.
The canonical fixed-temperature simulations 
at low temperatures tend to get
trapped in a few of a huge number of local-minimum-energy states
which are separated by high energy barriers.
One way to overcome this multiple-minima
problem is to perform a simulated annealing (SA) simulation \cite{SA},
and it has been widely used in biomolecular systems 
(see, e.g., Refs.~\cite{SA1}--\cite{KONF2} for earlier
applications).
The SA simulation mimicks the crystal-making process, 
and temperature is lowered very slowly from a sufficiently
high temperature to a low one during the SA simulation.
The Boltzmann weight factor is dynamically changed, and
so the thermal equilibrium is continuously broken.
Hence, although the global-minimum potential energy or
the value close to it may be found, accurate thermodynamic 
averages for fixed temperatures cannot be obtained.

A class of simulation methods, which are referred to as
the {\it generalized-ensemble algorithms}\index{generalized-ensemble algorithm},  
overcome both above difficulties, namely the multipole-minima
problem and inaccurate thermodynamic averages
(for reviews see, e.g., Refs.~\cite{RevHO1}--\cite{RevO09}).
In the generalized-ensemble algorithm, 
each state is weighted by an artificial,
non-Boltzmann probability
weight factor so that
a random walk\index{random walk} in potential energy space may be realized.
The random walk allows the simulation to escape from any
energy barrier and to sample much wider conformational
space than by conventional methods.
Unlike SA simulations, the weight factors are fixed
during the simulations so that the eventual reach to the 
thermal equilibrium is guaranteed.
From a single simulation run, one can obtain 
accurate ensemble
averages as functions of temperature 
by the single-histogram \cite{FS1}
and/or multiple-histogram \cite{FS2,WHAM} reweighting 
techniques\index{reweighting techniques}
(an extension of the multiple-histogram method is also referred to as the
{\it weighted histogram analysis method} (WHAM)\index{WHAM} \cite{WHAM}).

One of the most well-known generalized-ensemble algorithms is perhaps
the {\it multicanonical algorithm}\index{multicanonical algorithm} 
(MUCA\index{MUCA}) \cite{MUCA1,MUCA2}
(for reviews see, e.g., Refs.~\cite{BBook,RevJanke}).
The method is also referred to as {\it entropic sampling}
\cite{Lee,HS} and {\it adaptive umbrella sampling} \cite{MZ} 
{\it of the potential energy} \cite{BK}.
MUCA can also be considered as a sophisticated, ideal realization of
a class of algorithms called {\it umbrella sampling}\index{umbrella sampling}
\cite{US}.  
Also closely related methods are 
{\it Wang-Landau method}\index{Wang-Landau method}
\cite{Landau1,Landau2}, which is also 
referred to as {\it density of states Monte Carlo} \cite{dePablo},
and {\it Meta Dynamics} \cite{Meta}.
See also Ref.~\cite{THT04}.
MUCA and its generalizations have been applied to
spin systems
(see, e.g., Refs.~\cite{MUCA3}--\cite{BMO07}).
MUCA was also introduced to the molecular simulation field
\cite{HO}.
Since then MUCA and its generalizations have been extensively
used in many applications in
protein\index{protein} and other biomolecular 
systems \cite{HO94}--\cite{IO07}.
Molecular dynamics version of MUCA
has also been developed \cite{HOE96,NNK,BK} (see also
Refs.~\cite{Muna,HOE96} for the Langevin dynamics version).
MUCA has been extended so that flat distributions in
other variables instead of potential energy may be 
obtained (see, e.g., Refs.~\cite{BeHaNe93,JK95,KPV,BK2,ICK,BNO03,IO04}). 
This can be considered as a special case of the
multidimensional (or, multivariable) extensions of MUCA,
where a multidimensional random walk in potential energy space
and in other variable space is realized
(see, e.g., Refs.~\cite{KPV,BK2,HNSKN,OO03,IO07}).
In this article, we just present one of such methods, namely,
the 
{\it multibaric-multithermal algorithm}
\index{multibaric-multithermal algorithm} (MUBATH)
where a two-dimensional random walk in both potential energy space and
volume space is realized \cite{OO03}--\cite{okokmd06}. 
  
While a simulation in multicanonical ensemble performs a free
1D random walk in potential energy space, that in
{\it simulated tempering}\index{simulated tempering} (ST)\index{ST} \cite{ST1,ST2} 
(the method is also referred to as the
{\it method of expanded ensemble}\index{expanded ensemble} \cite{ST1})
performs a free random walk in temperature space
(for a review, see, e.g., Ref.~\cite{STrev}).
This random walk, in turn,
induces a random walk in potential energy space and
allows the simulation to escape from
states of energy local minima.
ST and its generalizations have
also been applied to chemical physics field and 
biomolecular systems \cite{EdP,IRB1,HO96a,HO96b,IRB2,MO4,STREM,PP07,ZCY09,ZM10,KS10}.

MUCA and ST are powerful, but the probability
weight factors are not {\it a priori} known and have to be
determined by iterations of short trial simulations.
This process can be non-trivial and very tedius for
complex systems with many degreees of freedom.

In the {\it replica-exchange method}\index{replica-exchange method} (REM)\index{REM}
\cite{RE1}--\cite{RE2}, the difficulty of weight factor
determination is greatly alleviated.  (A closely related method 
was independently developed in Ref.~\cite{RE3}.
Similar methods in which the same equations are used but
emphasis is laid on optimizations have
been developed \cite{KT,JWK}. 
REM is also referred to as 
{\it multiple Markov chain method}\index{multiple Markov chain method} \cite{RE4}
and {\it parallel tempering}\index{parallel tempering} \cite{STrev}.  Details
of literature
about REM and related algorithms can be found in
recent reviews \cite{IBArev,RevMSO,RevSMO}.)
In this method, a number of
non-interacting copies (or, replicas) of the original system
at different temperatures are
simulated independently and
simultaneously by the conventional MC or MD method. Every few steps,
pairs of replicas are exchanged with a specified transition
probability.
The weight factor is just the
product of Boltzmann factors, and so it is essentially known.

REM has already been used in many applications in protein 
systems \cite{H97}--\cite{kokubo-okamoto2}. 
Other molecular simulation fields have also been studied by
this method in various ensembles \cite{FD}--\cite{OKOM}.
Moreover, REM was introduced to the
quantum chemistry field \cite{ISNO}.
The details of molecular dynamics algorithm for REM,
which is referred to as the 
{\it Replica-Exchange Molecular Dynamics} (REMD)
have been worked out
in Ref.~\cite{SO}, and this
led to a wide application of REM
in the protein folding and related
problems (see, e.g., Refs.~\cite{Gar}--\cite{SD10}).

However, REM also has a computational difficulty:
As the number of degrees of freedom of the system increases,
the required number of replicas also greatly increases, whereas 
only a single replica is simulated in MUCA and ST.
This demands a lot of computer power for complex systems.
Our solution to this problem is: Use REM for the weight
factor determinations of MUCA, which is much
simpler than previous iterative methods of weight
determinations, and then perform a long MUCA 
production run.
The method is referred to as
the {\it replica-exchange multicanonical algorithm}\index{replica-exchange
multicanonical algorithm} (REMUCA)\index{REMUCA}
\cite{SO3,MSO03,MSO03b}.
In REMUCA,
a short replica-exchange simulation is performed, and the 
multicanonical weight factor is determined by
the multiple-histogram reweighting techniques \cite{FS2,WHAM}.
Another example of a combination of REM and ST is the
{\it replica-exchange simulated tempering}
\index{replica-exchange simulated tempering} 
(REST)\index{REST} \cite{MO4}.
In REST, a short replica-exchange simulation is performed, and
the simulated tempering weight factor is determined by
the multiple-histogram reweighting techniques \cite{FS2,WHAM}.

We have introduced two further extensions of REM,
which we refer to as 
{\it multicanonical replica-exchange method}
\index{multicanonical replica-exchange method}
(MUCAREM)\index{MUCAREM} \cite{SO3,MSO03,MSO03b} 
(see also Refs.~\cite{XB00,FYD02}) 
and {\it simulated tempering replica-exchange method}
\index{simulated tempering replica-exchange method}
(STREM)\index{STREM} \cite{STREM} (see also 
Ref.~\cite{FE03b} for a similar idea).
In MUCAREM, a replica-exchange simulation is
performed
with a small number of replicas each in multicanonical ensemble
of different energy ranges.  In STREM, on the other hand,
a replica-exchange simulation is performed with a small 
number of replicas in ``simulated tempering'' ensemble
of different temperature ranges.

Finally, one is naturally led to a multidimensional 
(or, multivariable) extension of REM, which we refer
to as the 
{\it multidimensional replica-exhcange method}
\index{multidimensional replica-exchange method} (MREM)\index{MREM}
\cite{SKO}
(see also Refs.~\cite{Huk2,YP}).  
(The method is also referred to as 
{\it generalized parallel sampling}\index{generalized parallel sampling}
\cite{WBS02},
{\it Hamiltonian replica-exchange method} 
\index{Hamiltonian replica-exchange method}\cite{FWT02}, 
and {\it Model Hopping}\index{Model Hopping}\cite{KH05}.)
Some other special cases of MREM can be found in, e.g., Refs. 
\cite{Dunw,FE03,LKFB05,ATdI06,LC06,Mu09,IOO10}.
Another special realization of MREM is 
{\it replica-exchange umbrella sampling}\index{replica-exchange umbrella sampling} 
(REUS)\index{REUS}
\cite{SKO} and it is particularly useful in free energy calculations
(see also Ref.~\cite{WEK03} for a similar idea).
In this article, we just present one of such methods, namely,
the replica-exchange method in the isobaric-isothermal ensemble,
where not only temperature values but also pressure values
are exchanged in the replica-exchange processes
\cite{NOSMO,OKOM,RevSO,PG,PGG}. (The results of the first such
application of the two-dimensional replica-exchange simulations
in the isobaric-isothermal ensemble were presented in Ref.~\cite{RevSO}.)

In this article, we describe the generalized-ensemble algorithms
mentioned above.  Namely, we first review the three familiar methods:
REM, ST, and MUCA.  We then describe various extensions 
of these methods \cite{SKO,MO09a,MO09b,M09,MO-10b}.
Examples of the results by some of these algorithms are then presented.

\section{GENERALIZED-ENSEMBLE ALGORITHMS}

\subsection{Replica-Exchange Method}

Let us consider a system of $N$ atoms of 
mass $m_k$ ($k=1, \cdots, N$)
with their coordinate vectors and
momentum vectors denoted by 
$q \equiv \{{\biq}_1, \cdots, {\biq}_N\}$ and 
$p \equiv \{{\bip}_1, \cdots, {\bip}_N\}$,
respectively.
The Hamiltonian $H(q,p)$ of the system is the sum of the
kinetic energy $K(p)$ and the potential energy $E(q)$:
\begin{equation}
H(q,p) =  K(p) + E(q)~,
\label{eqn1}
\end{equation}
where
\begin{equation}
K(p) =  \sum_{k=1}^N \frac{{\bip_k}^2}{2 m_k}~.
\label{eqn2}
\end{equation}

In the canonical ensemble at temperature $T$ 
each state $x \equiv (q,p)$ with the Hamiltonian $H(q,p)$
is weighted by the Boltzmann factor:
\begin{equation}
W_{\rm B}(x;T) = \exp \left(-\beta H(q,p) \right)~,
\label{eqn3}
\end{equation}
where the inverse temperature $\beta$ is defined by 
$\beta = 1/k_{\rm B} T$ ($k_{\rm B}$ is the Boltzmann constant). 
The average kinetic energy at temperature $T$ is then given by
\begin{equation}
\left< ~K(p)~ \right>_T =  
\left< \sum_{k=1}^N \frac{{\bip_k}^2}{2 m_k} \right>_T 
= \frac{3}{2} N k_{\rm B} T~.
\label{eqn4}
\end{equation}

Because the coordinates $q$ and momenta $p$ are decoupled
in Eq.~(\ref{eqn1}), we can suppress the kinetic energy
part and can write the Boltzmann factor as
\begin{equation}
W_{\rm B}(x;T) = W_{\rm B}(E;T) = \exp (-\beta E)~.
\label{eqn4b}
\end{equation}
The canonical probability distribution of potential energy
$P_{\rm NVT}(E;T)$ is then given by the product of the
density of states $n(E)$ and the Boltzmann weight factor
$W_{\rm B}(E;T)$:
\begin{equation}
 P_{\rm NVT}(E;T) \propto n(E) W_{\rm B}(E;T)~.
\label{eqn4c}
\end{equation}
Because $n(E)$ is a rapidly
increasing function and the Boltzmann factor decreases exponentially,
the canonical ensemble yields a bell-shaped distribution
of potential energy 
which has a maximum around the average energy at temperature $T$. 
The
conventional MC or MD simulations at constant temperature
are expected to yield $P_{\rm NVT}(E;T)$.
A MC simulation based on the Metropolis 
algorithm \cite{Metro} is performed with the following 
transition probability from a state $x$
of potential energy $E$ to a state $x^{\prime}$ of 
potential energy $E^{\prime}$:
\begin{equation}
w(x \rightarrow x^{\prime})
= {\rm min} \left(1,\frac{W_{\rm B}(E^{\prime};T)}{W_{\rm B}(E;T)}\right)
= {\rm min} \left(1,\exp \left(-\beta \Delta E \right)\right)~.
\label{eqn4d}
\end{equation}
where
\begin{equation}
\Delta E = E^{\prime} - E~.
\label{eqn4dp}
\end{equation}
A MD simulation, on the other hand, is based on the 
following Newton equations of motion:
\begin{eqnarray}
\dot{\bs{q}_k} &=& \frac{\bs{p}_k}{m_k}~, \\
\dot{\bs{p}_k} &=& 
- \frac{\partial E}{\partial \bs{q}_k} = \bs{f}_k~, 
\label{eqn4e}
\end{eqnarray}
where $\bs{f}_k$ is the force acting on the $k$-th atom
($k = 1, \cdots, N$).
This set of equations actually yield the microcanonical ensemble, however,
and we have to add a thermostat 
in order to obtain the canonical ensemble at temperature $T$.
Here, we just follow Nos{\'e}'s prescription~\cite{nose84,nosejcp84}, 
and we have
\begin{eqnarray}
\dot{\bs{q}}_k ~&=&~ \frac{\bs{p}_k}{m_k}~, \label{eqn4f1}\\
\dot{\bs{p}}_k ~&=&~ 
- \frac{\partial E}{\partial \bs{q}_k} 
- \frac{\dot{s}}{s}~\bs{p}_k~
= \bs{f}_k - \frac{\dot{s}}{s}~\bs{p}_k~, \label{eqn4f2}\\
\dot{s} ~&=&~ s~\frac{P_s}{Q}~, \label{eqn4f3}\\
\dot{P}_s ~&=&~ 
\sum_{k=1}^{N} \frac{{\bs{p}_k}^{2}}{m_k} - 3N k_{\rm B} T 
~=~3Nk_{\rm B} \left( T(t) - T \right)~, 
\label{eqn4f}
\end{eqnarray}
where $s$ is Nos{\'e}'s scaling parameter, 
$P_s$ is its conjugate momentum, $Q$ is its mass, and
the ``instantaneous temperature'' $T(t)$ is defined by
\begin{equation}
T(t) = \frac{1}{3N k_{\rm B}} \sum_{k=1}^{N} \frac{\bs{p}_k(t)^{2}}{m_k} ~.
\label{eqn4g}
\end{equation}
  
However, in practice, it is very difficult to obtain accurate canonical
distributions of complex systems at low temperatures by
conventional MC or MD simulation methods.
This is because simulations at low temperatures tend to get
trapped in one or a few of local-minimum-energy states.
This difficulty is overcome by, for instance, the generalized-ensemble
algorithms, which greatly enhance conformational sampling.

%%%%%%%%%%%%%%%%%%%%%%%%%%%%%%%%%%%%%%%%%%%%%%%%%%%%%%%%%%%%%%%%%%%%%
%\subsection{Replica-Exchange Method}
The {\it replica-exchange method} (REM) is one of effective generalized-ensemble
algorithms.
The system for REM consists of 
$M$ {\it non-interacting} copies (or, replicas) 
of the original system in the canonical ensemble
at $M$ different temperatures $T_m$ ($m=1, \cdots, M$).
We arrange the replicas so that there is always
exactly one replica at each temperature.
Then there exists a one-to-one correspondence between replicas
and temperatures; the label $i$ ($i=1, \cdots, M$) for replicas 
is a permutation of 
the label $m$ ($m=1, \cdots, M$) for temperatures,
and vice versa:
\begin{equation}
\left\{
\begin{array}{rl}
i &=~ i(m) ~\equiv~ f(m)~, \cr
m &=~ m(i) ~\equiv~ f^{-1}(i)~,
\end{array}
\right.
\label{eq4b}
\end{equation}
where $f(m)$ is a permutation function of $m$ and
$f^{-1}(i)$ is its inverse.

Let $X = \left\{x_1^{[i(1)]}, \cdots, x_M^{[i(M)]}\right\} 
= \left\{x_{m(1)}^{[1]}, \cdots, x_{m(M)}^{[M]}\right\}$ 
stand for a ``state'' in this generalized ensemble.
Each ``substate'' $x_m^{[i]}$ is specified by the
coordinates $q^{[i]}$ and momenta $p^{[i]}$
of $N$ atoms in replica $i$ at temperature $T_m$:
\begin{equation}
x_m^{[i]} \equiv \left(q^{[i]},p^{[i]}\right)_m~.
\label{eq5}
\end{equation}

Because the replicas are non-interacting, the weight factor for
the state $X$ in
this generalized ensemble is given by
the product of Boltzmann factors for each replica (or at each
temperature):
\begin{equation}
\begin{array}{rl}
W_{\rm REM}(X) 
&= \displaystyle{ \prod_{i=1}^{M}
\exp \left\{- \beta_{m(i)} 
H\left(q^{[i]},p^{[i]}\right) \right\} } 
 = \displaystyle{ \prod_{m=1}^{M}
  \exp \left\{- \beta_m 
H\left(q^{[i(m)]},p^{[i(m)]}\right)
 \right\} }~, \cr
&= \exp \left\{- \dis{\sum_{i=1}^M \beta_{m(i)} 
H\left(q^{[i]},p^{[i]}\right) } \right\}
 = \exp \left\{- \dis{\sum_{m=1}^M \beta_m 
H\left(q^{[i(m)]},p^{[i(m)]}\right) }
 \right\}~,
\end{array}
\label{eq7}
\end{equation}
where $i(m)$ and $m(i)$ are the permutation functions in 
Eq.~(\ref{eq4b}).

We now consider exchanging a pair of replicas in this
ensemble.  Suppose we exchange replicas $i$ and $j$ which are
at temperatures $T_m$ and $T_n$, respectively:  
\begin{equation}
X = \left\{\cdots, x_m^{[i]}, \cdots, x_n^{[j]}, \cdots \right\} 
\longrightarrow \ 
X^{\prime} = \left\{\cdots, x_m^{[j] \prime}, \cdots, x_n^{[i] \prime}, 
\cdots \right\}~. 
\label{eq8}
\end{equation}
Here, $i$, $j$, $m$, and $n$ are related by the permutation
functions in Eq.~(\ref{eq4b}),
and the exchange of replicas introduces a new 
permutation function $f^{\prime}$:
\begin{equation}
\left\{
\begin{array}{rl}
i &= f(m) \longrightarrow j=f^{\prime}(m)~, \cr
j &= f(n) \longrightarrow i=f^{\prime}(n)~. \cr
\end{array}
\right.
\label{eq8c}
\end{equation}

The exchange of replicas can be written in more detail as
\begin{equation}
\left\{
\begin{array}{rl}
x_m^{[i]} \equiv \left(q^{[i]},p^{[i]}\right)_m & \longrightarrow \ 
x_m^{[j] \prime} \equiv \left(q^{[j]},p^{[j] \prime}\right)_m~, \cr
x_n^{[j]} \equiv \left(q^{[j]},p^{[j]}\right)_n & \longrightarrow \ 
x_n^{[i] \prime} \equiv \left(q^{[i]},p^{[i] \prime}\right)_n~,
\end{array}
\right.
\label{eq9}
\end{equation}
where the definitions for $p^{[i] \prime}$ and $p^{[j] \prime}$
will be given below.
We remark that this process is equivalent to exchanging
a pair of temperatures $T_m$ and $T_n$ for the
corresponding replicas $i$ and $j$ as follows:
\begin{equation}
\left\{
\begin{array}{rl}
x_m^{[i]} \equiv \left(q^{[i]},p^{[i]}\right)_m & \longrightarrow \ 
x_n^{[i] \prime} \equiv \left(q^{[i]},p^{[i] \prime}\right)_n~, \cr
x_n^{[j]} \equiv \left(q^{[j]},p^{[j]}\right)_n & \longrightarrow \ 
x_m^{[j] \prime} \equiv \left(q^{[j]},p^{[j] \prime}\right)_m~.
\end{array}
\right.
\label{eq10}
\end{equation}

In the original implementation of the 
{\it replica-exchange method} (REM) \cite{RE1}--\cite{RE2},
Monte Carlo algorithm was used, and only the coordinates $q$
(and the potential energy
function $E(q)$)
had to be taken into account.  
In molecular dynamics algorithm, on the other hand, we also have to
deal with the momenta $p$.
We proposed the following momentum 
assignment in Eq.~(\ref{eq9}) (and in Eq.~(\ref{eq10})) \cite{SO}:
\begin{equation}
\left\{
\begin{array}{rl}
p^{[i] \prime} & \equiv \dis{\sqrt{\frac{T_n}{T_m}}} ~p^{[i]}~, \cr
p^{[j] \prime} & \equiv \dis{\sqrt{\frac{T_m}{T_n}}} ~p^{[j]}~,
\end{array}
\right.
\label{eq11}
\end{equation}
which we believe is the simplest and the most natural.
This assignment means that we just rescale uniformly 
the velocities of all the atoms 
in the replicas by
the square root of the ratio of the two temperatures so that
the temperature condition in Eq.~(\ref{eqn4}) may be satisfied
immediately after replica exchange is accepted.

The transition probability of this replica-exchange
process is given by the usual Metropolis criterion:
\begin{equation}
w(X \rightarrow X^{\prime}) \equiv
w\left( x_m^{[i]} ~\left|~ x_n^{[j]} \right. \right) 
= {\rm min}\left(1,\frac{W_{\rm REM}(X^{\prime})}
{W_{\rm REM}(X)}\right)
= {\rm min}\left(1,\exp \left( - \Delta \right)\right)~,
\label{eq15}
\end{equation}
where in the second expression 
(i.e., $w( x_m^{[i]} | x_n^{[j]} )$) 
we explicitly wrote the
pair of replicas (and temperatures) to be exchanged.
From Eqs.~(\ref{eqn1}), (\ref{eqn2}), (\ref{eq7}), and (\ref{eq11}), 
we have
\begin{equation}
\begin{array}{rl}
\dis{\frac{W_{\rm REM}(X^{\prime})}{W_{\rm REM}(X)}}
&= \exp \left\{ 
- \beta_m \left[K\left(p^{[j] \prime}\right) + E\left(q^{[j]}\right)\right] 
- \beta_n \left[K\left(p^{[i] \prime}\right) + E\left(q^{[i]}\right)\right]
\right. \cr
& \ \ \ \ \ \ \ \ \ \  \left.
+ \beta_m \left[K\left(p^{[i]}\right) + E\left(q^{[i]}\right)\right] 
+ \beta_n \left[K\left(p^{[j]}\right) + E\left(q^{[j]}\right)\right]
\right\}~, \cr
&= \exp \left\{ 
- \beta_m \dis{\frac{T_m}{T_n}} K\left(p^{[j]}\right)
- \beta_n \dis{\frac{T_n}{T_m}} K\left(p^{[i]}\right)
+ \beta_m K\left(p^{[i]}\right)
+ \beta_n K\left(p^{[j]}\right)
\right. \cr
& \ \ \ \ \ \ \ \ \ \  \left.
- \beta_m \left[E\left(q^{[j]}\right)
                - E\left(q^{[i]}\right)\right] 
- \beta_n \left[E\left(q^{[i]}\right)
                - E\left(q^{[j]}\right)\right] 
\right\}~.
\end{array}
\label{eq13}
\end{equation}
Because the kinetic energy terms in this equation
all cancel out, $\Delta$
in Eq.~(\ref{eq15}) becomes
\begin{eqnarray}
\Delta &=& \beta_m 
\left(E\left(q^{[j]}\right) - E\left(q^{[i]}\right)\right) 
- \beta_n
\left(E\left(q^{[j]}\right) - E\left(q^{[i]}\right)\right)~,
\label{eqn14a} \\
  &=& \left(\beta_m - \beta_n \right)
\left(E\left(q^{[j]}\right) - E\left(q^{[i]}\right)\right)~. 
\label{eqn14b}
\end{eqnarray}
Here, $i$, $j$, $m$, and $n$ are related by the permutation
functions in Eq.~(\ref{eq4b}) before the replica exchange:
\begin{equation}
\left\{
\begin{array}{ll}
i &= f(m)~, \cr
j &= f(n)~.
\end{array}
\right.
\label{eq13b}
\end{equation}

Note that after introducing the momentum rescaling in Eq.~(\ref{eq11}),
we have the same Metropolis criterion for replica exchanges, i.e.,
Eqs.~(\ref{eq15}) and (\ref{eqn14b}), for both MC and MD versions.

Without loss of generality we can
assume $T_1 < T_2 < \cdots < T_M$.
The lowest temperature
$T_1$ should be sufficiently low so that the simulation can explore the
global-minimum-energy region, and
the highest temperature $T_M$ should be sufficiently high so that
no trapping in an energy-local-minimum state occurs. 
A simulation of the 
{\it replica-exchange method} (REM)
is then realized by alternately performing the following two
steps:
\begin{enumerate}
\item Each replica in canonical ensemble of the fixed temperature 
is simulated $simultaneously$ and $independently$
for a certain MC or MD steps. 
\item A pair of replicas at neighboring temperatures,
say $x_m^{[i]}$ and $x_{m+1}^{[j]}$, are exchanged
with the probability
$w\left( x_m^{[i]} ~\left|~ x_{m+1}^{[j]} \right. \right)$ 
in Eq.~(\ref{eq15}).
\end{enumerate}
Note that in Step 2 we exchange only pairs of replicas corresponding to
neighboring temperatures, because
the acceptance ratio of the exchange process decreases exponentially
with the difference of the two $\beta$'s (see Eqs.~(\ref{eqn14b})
and (\ref{eq15})).
Note also that whenever a replica exchange is accepted
in Step 2, the permutation functions in Eq.~(\ref{eq4b})
are updated.  A random walk in ``temperature space'' is
realized for each replica, which in turn induces a random
walk in potential energy space.  This alleviates the problem
of getting trapped in states of energy local minima.

The REM simulation is particularly suitable for parallel
computers.  Because one can minimize the amount of information
exchanged among nodes, it is best to assign each replica to
each node (exchanging pairs of temperature values among nodes
is much faster than exchanging coordinates and momenta).
This means that we keep track of the permutation function
$m(i;t)=f^{-1}(i;t)$ in Eq.~(\ref{eq4b}) as a function
of MC or MD step $t$ during the simulation.
After parallel canonical MC or MD simulations for a certain
steps (Step 1), $M/2$ pairs of
replicas corresponding to neighboring temperatures
are simulateneously exchanged (Step 2), and the pairing is alternated 
between the two possible choices, i.e., ($T_1,T_2$), ($T_3,T_4$), $\cdots$
and ($T_2,T_3$), ($T_4,T_5$), $\cdots$.
  
After a long production run of a replica-exchange simulation, 
the canonical expectation value of a physical quantity $A$
at temperature $T_m$ ($m=1, \cdots, M$) can be calculated
by the usual arithmetic mean:
\begin{equation}
<A>_{T_m} = \frac{1}{n_{m}} \sum_{k=1}^{n_{m}}
A\left(x_{m}(k)\right)~,
\label{Eqn7}
\end{equation}
where $x_m(k)$ ($k=1,\cdots,n_m$) are the configurations 
obtained at temperature $T_m$
and $n_{m}$ is the total number of measurements made
at $T=T_m$.
The expectation value at any intermediate temperature
$T$ ($= 1/k_{\rm B} \beta$)
can also be obtained as follows:
\begin{equation} 
<A>_T =  \frac{ \displaystyle{ 
\sum_E~ A (E) P_{\rm NVT}(E;T)} }
{\displaystyle{ \sum_E~ P_{\rm NVT}(E;T)} } =  \frac{ \displaystyle{ 
\sum_E~ A (E) n(E) \exp(-\beta E) } }
{\displaystyle{ \sum_E~ n(E) \exp(-\beta E) } }~. 
\label{eqn10a}
\end{equation}
The summation instead of integration is used
in Eq.~(\ref{eqn10a}), because we often
discretize the potential energy $E$ with step size $\epsilon$
($E = E_i; i=1, 2, \cdots$).
Here, the explicit form of the physical quantity $A$
should be known as a function of potential energy $E$.
For instance, $A(E)=E$ gives the average
potential energy $<E>_T$ as a function of temperature, and 
$A(E) = \beta^2 (E - <E>_T)^2$ gives specific heat.

The density of states $n(E)$ in Eq.~(\ref{eqn10a}) is given by
the multiple-histogram reweighting techniques \cite{FS2,WHAM}
as follows.
Let $N_m(E)$ and $n_m$ be respectively
the potential-energy histogram and the total number of
samples obtained at temperature $T_m=1/k_{\rm B} \beta_m$
($m=1, \cdots, M$). 
The best estimate of the density of states is then given by \cite{FS2,WHAM}
\begin{equation}
n(E) = \frac{\dis{\sum_{m=1}^M ~g_m^{-1}~N_m(E)}}
{\dis{\sum_{m=1}^M ~g_m^{-1}~n_m~\exp (f_m-\beta_m E)}}~,
\label{Eqn8a}
\end{equation}
where we have for each $m$ ($=1, \cdots, M$)
\begin{equation}
\exp (-f_m) = \sum_{E} ~n(E)~\exp (-\beta_m E)~.
\label{Eqn8b}
\end{equation}
Here, $g_m = 1 + 2 \tau_m$,
and $\tau_m$ is the integrated
autocorrelation time at temperature $T_m$.
For many systems the quantity $g_m $ can safely 
be set to be a constant in the reweighting 
formulae \cite{WHAM}, and hereafter we set $g_m =1$. 

Note that
Eqs.~(\ref{Eqn8a}) and
(\ref{Eqn8b}) are solved self-consistently
by iteration \cite{FS2,WHAM} to obtain
the density of states $n(E)$ and
the dimensionless Helmholtz free energy $f_m$.
Namely, we
can set all the $f_m$ ($m=1, \cdots, M$) to, e.g., zero initially.
We then use Eq.~(\ref{Eqn8a}) to obtain 
$n(E)$, which is substituted into
Eq.~(\ref{Eqn8b}) to obtain next values of $f_m$, and so on.

Moreover, the ensemble averages of any physical quantity $A$
(including those that cannot be expressed as functions
of potential energy) at any temperature 
$T$ ($= 1/k_{\rm B} \beta$) can now be obtained from
the ``trajectory'' of configurations of 
the production 
run.  Namely, we first obtain $f_m$ ($m=1, \cdots, M$) by solving
Eqs. (\ref{Eqn8a}) and (\ref{Eqn8b}) self-consistently, 
and then we have \cite{MSO03} (see also \cite{SC08})
\begin{equation}
<A>_T  =  \frac{ \displaystyle{ 
\sum_{m=1}^{M} \sum_{k=1}^{n_m} A(x_m(k))  \frac{1}
{\displaystyle{\sum_{\ell=1}^{M} n_{\ell} 
\exp \left[ f_{\ell} - \beta_{\ell} E(x_m(k)) \right]}}
\exp \left[ -\beta E(x_m(k)) \right] }}
{\displaystyle{ 
\sum_{m=1}^{M} \sum_{k=1}^{n_m} \frac{1}
{\displaystyle{\sum_{\ell=1}^{M} n_{\ell} 
\exp \left[ f_{\ell} - \beta_{\ell} E(x_m(k)) \right]}}
\exp \left[ -\beta E(x_m(k)) \right] }}~,
\label{eqn17}
\end{equation}
where $x_m(k)$ ($k=1,\cdots,n_m$) are the configurations 
obtained at temperature $T_m$. \\ 

Eqs. (\ref{eqn10a}) and (\ref{Eqn8a}) or any other equations
which involve summations of
exponential functions often encounter with numerical difficulties
such as overflows.  These can be overcome by using, 
for instance,
the following equation \cite{BBook,BergLog}:
For $C=A+B$ (with $A>0$ and $B>0$) we have
\begin{equation}
\begin{array}{rl} 
\ln C &= \ln \left[{\rm max}(A,B) \left(1 +
 \displaystyle{\frac{{\rm min}(A,B)}{{\rm max}(A,B)}} 
 \right) \right]~, \\
 &= {\rm max}(\ln A, \ln B) +
 \ln \left\{1+\exp \left[{\rm min}(\ln A,\ln B) -
 {\rm max}(\ln A,\ln B) \right] \right\}~.
\end{array}
\label{eqn10e}
\end{equation}
     
We now give more details about the momentum rescaling
in Eq.~(\ref{eq11}) \cite{MO-10a}.
Actually, Eq.~(\ref{eq11}) is only valid for
the Langevin dynamics \cite{Lang}, Andersen thermostat \cite{Andersen}, 
and Gaussian constraint method \cite{Gcons1,Gcons2,Gcons3}.
The former two thermostats are based on the
weight factor of Eq.~(\ref{eqn3}) with Eqs.~(\ref{eqn1}) and (\ref{eqn2}), 
and the Gaussian contraint method is based on the
following weight factor:
\begin{equation}
W(q,p) = \delta 
\left( \sum_{k=1}^N \frac{\bip_k^2}{2m_k} - \frac{gk_{\rm{B}}T}{2} \right) 
\exp \left[ -\beta E(q) \right]~. 
\end{equation}

For Nos\'{e}'s method \cite{nose84,nosejcp84},
which gives the equations of motion in
Eqs.~(\ref{eqn4f1})--(\ref{eqn4f}), 
the Nos\'{e} Hamiltonian is given by
\begin{equation}
H_{\rm Nose} = \sum_{k=1}^N\frac{\tilde{\bip}_k^2}{2m_k s^2} + E(q) + 
\frac{P_s^2}{2Q} + gk_{\rm{B}}T\log s .
\label{NoseH}
\end{equation}
Here, $g$ $(=3N)$ is the number of degrees of freedom, $s$ is a position 
variable of the thermostat, $P_s$ is a momentum conjugate to $s$, 
and $\tilde{\bip}_k$ is a virtual momentum, which is related to the real 
momenta $\bip_k$ as $\bip_k = \tilde{\bip}_k / s$.
The weight factor for the Nos\'{e}'s method is then given by
\begin{equation}
W(q,\tilde{p},s,P_s) = \delta \left( H_{\rm{Nose}} - \mathcal{E} \right)~,
\end{equation}
where $\mathcal{E}$ is the initial value of $H_{\rm{Nose}}$.
Namely, in the Nos\'{e}'s method, the entire system including the
thermostat is in the microcanonical ensemble.
Note that the mass $Q$ of the thermostat
can have different values in each replica in REMD simulations.
The rescaling method for the 
Nos\'{e} thermostat is given by Eq.~(\ref{eq11})
and
\begin{equation}
\label{mom_scale_np}
P_s^{[i]\prime} = \sqrt{\frac{T_nQ_n}{T_mQ_m}}P_s^{[i]}, \quad 
P_s^{[j]\prime} = \sqrt{\frac{T_mQ_m}{T_nQ_n}}P_s ^{[j]},
\end{equation}
\begin{equation}
	\label{np}
\begin{array}{rl}
	s^{[i]\prime} &= s^{[i]} \exp \left[\dis \frac{1}{gk_{\rm{B}}} \left( \frac{E(q^{[i]})-\mathcal{E}_m}{T_m} - \frac{E(q^{[i]})-\mathcal{E}_n}{T_n} \right) \right] , \\
	s^{[j]\prime} &= s^{[j]} \exp \left[\dis \frac{1}{gk_{\rm{B}}} \left( \frac{E(q^{[j]})-\mathcal{E}_n}{T_n} - \frac{E(q^{[j]})-\mathcal{E}_m}{T_m} \right) \right] ,
\end{array}
\end{equation}
where $\mathcal{E}_m$ and $\mathcal{E}_n$ are the initial values of 
$H_{\rm Nose}$ in the simulations with $T_m$ and $T_n$, respectively,
before the replica exchange.
Note that the real momenta have to be used in the rescaling method 
in Eq. (\ref{eq11}), not the virtual momenta.

For the Nos\'{e}-Hoover thermostat \cite{NoseHoover}, the states 
are specified by the following weight factor:
\begin{equation}
W(q,p,\zeta) = \exp \left[ -\beta \left( \sum_{k=1}^N 
\frac{\bip_k^2}{2m_k} + E(q) + \frac{Q}{2}\zeta^2 \right) \right],
\end{equation}
where $\zeta$ is a velocity of the thermostat and $Q$ is its mass parameter.
The rescaling method for the Nos\'{e}-Hoover thermostat
is given by Eq.~(\ref{eq11})
and 
\begin{equation}
\label{mom_scale_nh}
\zeta^{[i]\prime} = \sqrt{\frac{T_nQ_m}{T_mQ_n}}\zeta^{[i]}, \quad \zeta^{[j]\prime} = \sqrt{\frac{T_mQ_n}{T_nQ_m}}\zeta^{[j]},
\end{equation}
where $Q_m$ and $Q_n$ are the mass parameters in the replicas at temperature 
values $T_m$ and $T_n$, respectively, before the replica exchange. 

The rescaling method for the Nos\'{e}-Hoover thermostat can 
be generalized to the Nos\'{e}-Hoover chains~\cite{NoseHooverC} 
in a similar way.
The weight factor for
the Nos\'{e}-Hoover chains is given by
\begin{equation}
W(q,p,\zeta_1,\cdots ,\zeta_{\cL}) = 
\exp \left[ -\beta \left( \sum_{k=1}^N \frac{\bip_k^2}{2m_k} + E(q) 
+ \sum_{\ell=1}^{\cL}\frac{Q_{\ell}}{2}\zeta_{\ell}^2 \right) \right], 
\end{equation}
where $\cL$ is the number of thermostats, 
$\zeta_{\ell} \ (\ell =1,\cdots ,\cL)$ is the velocity of the $\ell$-th thermostat,
and $Q_{\ell} \ (\ell =1,\cdots ,\cL)$ is a mass parameter corresponding to 
the $\ell$-th thermostat.
A rescaling method for REMD with the Nos\'{e}-Hoover chains is given 
by Eq.~(\ref{eq11}) and the following:
\begin{equation}
\label{mom_scale_nhc}
	\zeta_{\ell}^{[i]\prime} = 
\sqrt{\frac{T_nQ_{m,\ell}}{T_mQ_{n,\ell}}}\zeta_{\ell}^{[i]}, \quad 
\zeta_{\ell}^{[j]\prime} = \sqrt{\frac{T_mQ_{n,\ell}}{T_nQ_{m,\ell}}}\zeta_{\ell}^{[j]}, 
\ (\ell=1,\cdots ,\cL),
\end{equation}
where $Q_{m,\ell}$ and $Q_{n,\ell}$ are the mass parameters in the replicas 
at temperature values $T_m$ and $T_n$, respectively, which correspond to 
the $\ell$-th thermostat. 

%\subsection{Nos\'{e}-Poincar\'{e} thermostat}
In the Nos\'{e}-Poincar\'{e} thermostat \cite{NoseP}, the states are specified 
by $x \equiv (q,\tilde{p},s,P_s)$ and the weight factor
is given by 
\begin{equation}
W(q,\tilde{p},s,P_s )  \propto \delta \left[ s \left( H_{\rm{Nose}} - \mathcal{E} \right) \right],
\end{equation}
where $H_{\rm Nose}$ is 
the Nos\'{e} Hamiltonian in Eq.~(\ref{NoseH})
and $\mathcal{E}$ is its initial value.
A rescaling method of the Nos\'{e}-Poincar\'{e} thermostat is the same as in
the Nos\'{e}'s thermostat and given by 
Eqs.~(\ref{eq11}), (\ref{mom_scale_np}), and (\ref{np}) above.

\subsection{Simulated Tempering}
We now introduce another generalized-ensemble algorithm, the {\it simulated tempering} (ST)
method \cite{ST1,ST2}.  In this method temperature itself becomes a
dynamical variable, and both the configuration and the temperature are updated
during the simulation with a weight:
\begin{equation}
W_{\rm ST} (E;T) = \exp \left(-\beta E + a(T) \right)~,
\label{Eqn1}
\end{equation}
where the function $a(T)$ is chosen so that the probability distribution
of temperature is flat:
\begin{equation}
P_{\rm ST}(T) = \int dE~ n(E)~ W_{{\rm ST}} (E;T) =
\int dE~ n(E)~ \exp \left(-\beta E + a(T) \right) = {\rm constant}~.
\lab{Eqn2}
\end{equation}
Hence, in simulated tempering, {\it temperature} is sampled
uniformly. A free random walk in temperature space
is realized, which in turn
induces a random walk in potential energy space and
allows the simulation to escape from
states of energy local minima.

In the numerical work we discretize the temperature in
$M$ different values, $T_m$ ($m=1, \cdots, M$).  Without loss of
generality we can order the temperature
so that $T_1 < T_2 < \cdots < T_M$.  
The probability weight factor in Eq.~(\ref{Eqn1}) is now written as
\begin{equation}
W_{\rm ST}(E;T_m) = \exp (-\beta_m E + a_m)~,
\label{Eqn3}
\end{equation}
where $a_m=a(T_m)$ ($m=1, \cdots, M$).
Note that from Eqs.~(\ref{Eqn2}) and (\ref{Eqn3}) we have
\begin{equation}
\exp (-a_m) \propto \int dE~ n(E)~ \exp (- \beta_m E)~.
\label{Eqn4}
\end{equation}
The parameters $a_m$ are therefore ``dimensionless'' Helmholtz free energy
at temperature $T_m$
(i.e., the inverse temperature $\beta_m$ multiplied by
the Helmholtz free energy).

Once the parameters $a_m$ are determined and the initial configuration and
the initial temperature $T_m$ are chosen,
a simulated tempering simulation is realized by alternately
performing the following two steps \cite{ST1,ST2}:
\begin{enumerate}
\item A canonical MC or MD simulation at the fixed temperature $T_m$
is carried out for a certain steps.
\item The temperature $T_m$ is updated to the neighboring values
$T_{m \pm 1}$ with the configuration fixed.  The transition probability of
this temperature-updating
process is given by the following Metropolis criterion (see Eq.~(\ref{Eqn3})):
\begin{equation}
w(T_m \rightarrow T_{m \pm 1})
= {\rm min}\left(1,\frac{W_{\rm ST}(E;T_{m \pm 1})}{W_{\rm ST}(E;T_m)}\right)
= {\rm min}\left(1,\exp \left( - \Delta \right)\right)~,
\label{Eqn5}
\end{equation}
where
\begin{equation}
\Delta = \left(\beta_{m \pm 1} - \beta_m \right) E
- \left(a_{m \pm 1} - a_m \right)~.
\label{Eqn6}
\end{equation}
\end{enumerate}
Note that in Step 2 we update the temperature only to
the neighboring temperatures in order to secure sufficiently
large acceptance ratio of temperature updates.

We remark that when MD simulations are performed in Step 1 above, 
we also have to deal with the momenta $p$,
and the kinetic energy term should be included in the weight factor. 
When temperature $T_{m_0 \pm 1}$ is accepted for $T$-update in Step 2,  
we rescale the momenta in the same way as in REMD \cite{SO,MO09b,MO-10b}: 
\begin{equation}
\dis {{{\bs p}_k}^{\prime}}=\sqrt{ \frac{T_{m_0 \pm 1}}{T_{m_0}} } ~{\bs p}_k~.  
\end{equation}
The kinetic energy terms then cancel out in Eq.~(\ref{Eqn6}) 
and we can use the same $\Delta$ 
in the Metropolis criterion in Step 2 for both MC and MD simulations.   
Similar momentum scaling formulae given above
should also be introduced for various other thermostats \cite{MO-10a}.
      
The simulated tempering
parameters $a_m=a(T_m)$ ($m=1, \cdots, M$)
can be determined by iterations of short trial simulations
(see, e.g.,  Refs.~\cite{STrev,IRB1,HO96b} for details).
This process can be non-trivial and very tedius for complex
systems.

After the optimal simulated tempering weight factor is determined,
one performs a long simulated tempering run once.
The canonical expectation value of a physical quantity $A$
at temperature $T_m$ ($m=1, \cdots, M$) can be calculated
by the usual arithmetic mean 
from Eq.~(\ref{Eqn7}).
The expectation value at any intermediate temperature
can also be obtained from
Eq.~(\ref{eqn10a}), where the 
density of states is given by
the multiple-histogram reweighting techniques \cite{FS2,WHAM}.
Namely, let $N_m(E)$ and $n_m$ be respectively
the potential-energy histogram and the total number of
samples obtained at temperature $T_m=1/k_{\rm B} \beta_m$
($m=1, \cdots, M$). 
The best estimate of the density of states is then given by 
solving Eqs.~(\ref{Eqn8a}) and (\ref{Eqn8b}) self-consistently.

Moreover, the ensemble averages of any physical quantity $A$
(including those that cannot be expressed as functions
of potential energy) at any temperature 
$T$ ($= 1/k_{\rm B} \beta$) can now be obtained from
Eq.~(\ref{eqn17}).
\subsection{Multicanonical Algorithm}
The third generalized-ensemble algorithm that we present
is the 
{\it multicanonical algorithm} (MUCA) \cite{MUCA1,MUCA2}.
In the multicanonical ensemble, each state is weighted by
a non-Boltzmann weight
factor $W_{\rm MUCA}(E)$ (which we refer to as the {\it multicanonical
weight factor}) so that a uniform potential energy
distribution $P_{\rm MUCA}(E)$ is obtained:
\begin{equation}
 P_{\rm MUCA}(E) \propto n(E) W_{\rm MUCA}(E) \equiv {\rm constant}~.
\label{eqn5}
\end{equation}
The flat distribution implies that
a free one-dimensional random walk in the potential energy space is realized
in this ensemble.
This allows the simulation to escape from any local minimum-energy states
and to sample the configurational space much more widely than 
the conventional canonical MC or MD methods.

The definition in Eq.~(\ref{eqn5}) implies that
the multicanonical
weight factor is inversely proportional to the density of
states, and we can write it as follows:
\begin{equation}
W_{\rm MUCA}(E) \equiv \exp \left[-\beta_a E_{\rm MUCA}(E;T_a) \right]
= \frac{1}{n(E)}~,
\label{eqn6}
\end{equation}
where we have chosen an arbitrary reference
temperature, $T_a = 1/k_{\rm B} \beta_a$, and
the ``{\it multicanonical potential energy}''
is defined by
\begin{equation}
 E_{\rm MUCA}(E;T_a) \equiv k_{\rm B} T_a \ln n(E) = T_a S(E)~.
\label{eqn7}
\end{equation}
Here, $S(E)$ is the entropy in the microcanonical
ensemble.
Because the density of states of the system is usually unknown,
the multicanonical weight factor
has to be determined numerically by iterations of short preliminary
runs \cite{MUCA1,MUCA2}. 

A multicanonical MC simulation
is performed, for instance, with the usual Metropolis criterion \cite{Metro}:
The transition probability of state $x$ with potential energy
$E$ to state $x^{\prime}$ with potential energy $E^{\prime}$ is given by
\begin{equation}
w(x \rightarrow x^{\prime})
= {\rm min} \left(1,\frac{W_{\rm MUCA}(E^{\prime})}{W_{\rm MUCA}(E)}\right)
= {\rm min} \left(1,\frac{n(E)}{n(E^{\prime})}\right)
= {\rm min} \left(1,\exp \left( - \beta_a \Delta E_{\rm MUCA} \right)\right)~,
\label{eqn8}
\end{equation}
where
\begin{equation}
\Delta E_{\rm MUCA} = E_{\rm MUCA}(E^{\prime};T_a) - E_{\rm MUCA}(E;T_a)~.
\label{eqn9}
\end{equation}
The MD algorithm in the multicanonical ensemble
also naturally follows from Eq.~(\ref{eqn6}), in which the
regular constant temperature MD simulation
(with $T=T_a$) is performed by replacing $E$ by $E_{\rm MUCA}$
in Eq.~(\ref{eqn4f2})~\cite{HOE96,NNK}:  
\begin{equation}
\dot{\bs{p}}_k ~=~ - \frac{\partial E_{\rm MUCA}(E;T_a)}{\partial \bs{q}_k}
- \frac{\dot{s}}{s}~\bs{p}_k
~=~ \frac{\partial E_{\rm MUCA}(E;T_a)}{\partial E}~\bs{f}_k
- \frac{\dot{s}}{s}~\bs{p}_k~.
\label{eqn9a}
\end{equation}

If the exact multicanonical weight factor $W_{\rm MUCA}(E)$ is known, 
one can calculate the ensemble averages of any physical quantity $A$ 
at any temperature 
$T$ ($= 1/k_{\rm B} \beta$) 
from Eq.~(\ref{eqn10a}), 
where the density of states is given by (see Eq. (\ref{eqn6}))
\begin{equation} 
n(E)= \frac{1}{W_{\rm MUCA}(E)}~.
\label{eqn10b}
\end{equation}

In general, the multicanonical weight factor $W_{\rm MUCA}(E)$, or
the density of states $n(E)$, is not 
$a \ priori$ known, and one needs its estimator 
for a numerical simulation. 
This estimator is usually obtained 
from iterations of short trial multicanonical simulations. 
The details of this process
are described, for instance, in Refs.~\cite{MUCA3,OH}.
However, the iterative process can be non-trivial and very tedius for
complex systems.

In practice, it is impossible to obtain the ideal
multicanonical weight factor with completely uniform
potential energy distribution.
The question is when to stop the iteration for the
weight factor determination.
Our criterion for a satisfactory weight factor is that 
as long as we do get a random walk in potential energy space,
the probability distribution $P_{\rm MUCA}(E)$ does not have to
be completely
flat with a tolerance of, say, an order of magnitude deviation.
In such a case, we usually 
perform with this weight 
factor a multicanonical simulation with high statistics
(production run) in order to get even better estimate
of the density of states. 
Let $N_{\rm MUCA}(E)$ be the histogram
of potential energy distribution 
$P_{\rm MUCA} (E)$ 
obtained by this production run.
The best estimate of the density of states can then be
given by the single-histogram reweighting
techniques \cite{FS1} as follows (see the proportionality
relation in Eq. (\ref{eqn5})):
\begin{equation} 
n(E)= \displaystyle{\frac{N_{\rm MUCA}(E)}{W_{\rm MUCA}(E)}}~.
\label{eqn10c}
\end{equation}
By substituting this quantity into Eq. (\ref{eqn10a}),
one can calculate ensemble averages of physical
quantity $A(E)$ as a function of temperature.
Moreover, the ensemble averages of any physical quantity $A$
(including those that cannot be expressed as functions
of potential energy) at any temperature 
$T$ ($= 1/k_{\rm B} \beta$) can also be obtained as long as
one stores the ``trajectory'' of configurations 
from the production run.  
Namely, we have
\begin{equation} 
<A>_T  =  \frac{ \displaystyle{ 
\sum_{k=1}^{n_s} A(x_k) W_{\rm MUCA}^{-1} (E(x_k))
\exp \left[-\beta E(x_k) \right] } }
{\displaystyle{ 
\sum_{k=1}^{n_s} W_{\rm MUCA}^{-1} (E(x_k)) 
\exp \left[-\beta E(x_k) \right] } }~, 
\label{eqn10d}
\end{equation}
where $x_k$ is the configuration at the
$k$-th MC (or MD) step and $n_s$ is the total number 
of configurations stored. 
Note that when $A$ is a function of $E$, Eq.~(\ref{eqn10d}) reduces to
Eq.~(\ref{eqn10a}) where the density of states is given by
Eq.~(\ref{eqn10c}).

Some remarks are in order.
The major advantage of REM over other generalized-ensemble
methods such as simulated tempering \cite{ST1,ST2}
and multicanonical algorithm \cite{MUCA1,MUCA2}
lies in the fact that the weight factor 
is {\it a priori} known (see Eq.~(\ref{eq7})), while
in simulated tempering and multicanonical algorithm the determination of the
weight factors can be very tedius and time-consuming.
In REM, however, the number of required replicas increases greatly
($\propto \sqrt{N}$) as the system size $N$ increases \cite{RE1}, 
while only one replica is used
in simulated tempering and multicanonical algorithm.
This demands a lot of computer power for complex systems. 
Moreover, so long as optimal weight factors can be obtained,
simulated tempering and multicanonical algorithm are more efficient 
in sampling than the replica-exchange method \cite{RevSMO,STREM,MSO03b,SO5}.
  
\subsection{Replica-Exchange Simulated Tempering and 
Replica-Exchange Multicanonical Algorithm}

The {\it replica-exchange simulated tempering} (REST) \cite{MO4}  
and the {\it replica-exchange multicanonical algorithm} (REMUCA) 
\cite{SO3,MSO03,MSO03b}
overcome
both the difficulties of ST and MUCA
(the weight factor
determinations are non-trivial)
and REM (many replicas, or a lot of computation time, are required).

In REST \cite{MO4}, 
we first perform a short REM simulation (with $M$ replicas)
to determine the simulated tempering
weight factor and then perform with this weight
factor a regular ST simulation with high statistics.
The first step is accomplished by 
the multiple-histogram reweighting
techniques \cite{FS2,WHAM}, which give
the dimensionless Helmholtz free energy.
Let $N_m(E)$ and $n_m$ be respectively
the potential-energy histogram and the total number of
samples obtained at temperature $T_m$ ($=1/k_{\rm B} \beta_m$) 
of the REM run.
The dimensionless Helmholtz free energy $f_m$ 
is then given by solving 
Eqs.~(\ref{Eqn8a}) and (\ref{Eqn8b}) self-consistently by iteration.

Once the estimate of the dimensionless Helmholtz free energy $f_m$ are
obtained, the simulated tempering 
weight factor can be directly determined by using
Eq.~(\ref{Eqn3}) where we set $a_m = f_m$ (compare Eq.~(\ref{Eqn4})
with Eq.~(\ref{Eqn8b})).
A long simulated tempering run is then performed with this
weight factor.  
Let $N_m(E)$ and $n_m$ be respectively
the potential-energy histogram and the total number of
samples obtained at temperature $T_m$ ($=1/k_{\rm B} \beta_m$) from this
simulated tempering run.  The multiple-histogram
reweighting techniques of Eqs.~(\ref{Eqn8a}) and (\ref{Eqn8b}) can be used
again to obtain the best estimate of the density of states
$n(E)$.
The expectation value of a physical quantity $A$
at any temperature $T~(= 1/k_{\rm B} \beta)$ is then calculated from
Eq.~(\ref{eqn10a}).

We now present the {\it replica-exchange multicanonical algorithm} (REMUCA) 
\cite{SO3,MSO03,MSO03b}.
In REMUCA, just as in REST, we first perform a short REM simulation (with $M$ replicas)
to determine the
multicanonical weight factor and then perform with this weight
factor a regular multicanonical simulation with high statistics.
The first step is accomplished by the multiple-histogram reweighting
techniques \cite{FS2,WHAM}, which give the density of states.
Let $N_m(E)$ and $n_m$ be respectively
the potential-energy histogram and the total number of
samples obtained at temperature $T_m$ ($=1/k_{\rm B} \beta_m$) 
of the REM run.
The density of states $n(E)$ is then given by solving 
Eqs.~(\ref{Eqn8a}) and (\ref{Eqn8b}) self-consistently by iteration.

Once the estimate of the density of states is obtained, the
multicanonical weight factor can be directly determined from
Eq.~(\ref{eqn6}) (see also Eq.~(\ref{eqn7})).
Actually, the density of states $n(E)$ and 
the multicanonical potential energy, $E_{\rm MUCA}(E;T_0)$,
thus determined are only reliable in the following range:
\begin{equation}
E_1 \le E \le E_M~,
\label{eqn29}
\end{equation}
where 
\begin{equation}
\left\{
\begin{array}{rl}
E_1 &=~ <E>_{T_1}~, \\
E_M &=~ <E>_{T_M}~,
\end{array}
\right.
\label{eqn29b}
\end{equation}
and $T_1$ and $T_M$ are respectively the lowest and the highest
temperatures used in the REM run.
Outside this range we extrapolate
the multicanonical potential energy linearly \cite{SO3}:
\begin{equation}
 {\cal E}_{\rm MUCA}(E) \equiv \left\{
   \begin{array}{@{\,}ll}
   \left. \dis{\frac{\partial E_{\rm MUCA}(E;T_0)}{\partial E}}
        \right|_{E=E_1} (E - E_1)
             + E_{\rm MUCA}(E_1;T_0)~, &
         \mbox{for $E < E_1$,} \\
         E_{\rm MUCA}(E;T_0)~, &
         \mbox{for $E_1 \le E \le E_M$,} \\
   \left. \dis{\frac{\partial E_{\rm MUCA}(E;T_0)}{\partial E}}
        \right|_{E=E_M} (E - E_M)
             + E_{\rm MUCA}(E_M;T_0)~, &
         \mbox{for $E > E_M$.}
   \end{array}
   \right.
\label{eqn31}
\end{equation}
     
For Monte Carlo method, 
the weight factor for REMUCA is given by 
substituting Eq.~(\ref{eqn31}) into Eq.~(\ref{eqn6})
\cite{SO3,MSO03}:
\begin{equation}
W_{\rm MUCA}(E) = \exp \left[-\beta_0 {\cal E}_{\rm MUCA}(E) \right]
 = \left\{
   \begin{array}{@{\,}ll}
   \dis{\exp \left(-\beta_1 E \right)}~, &
         \mbox{for $E < E_1$,} \\
   \dis{\frac{1}{n(E)}}~, &
         \mbox{for $E_1 \le E \le E_M$,} \\
   \dis{\exp \left(-\beta_M E \right)}~, &
         \mbox{for $E > E_M$.}
   \end{array}
   \right.
\label{eqn31d}
\end{equation}
    
The multicanonical MC and MD runs  
are then performed respectively with
the Metropolis criterion of Eq.~(\ref{eqn8})
and with the modified Newton equation 
in Eq.~(\ref{eqn9a}), 
in which 
${\cal E}_{\rm MUCA}(E)$ in
Eq.~(\ref{eqn31}) is substituted into $E_{\rm MUCA}(E;T_0)$.
We expect to obtain a flat potential energy distribution in
the range of Eq.~(\ref{eqn29}).
Finally, the results are analyzed by the single-histogram
reweighting techniques as described in Eq.~(\ref{eqn10c})
(and Eq.~(\ref{eqn10a})).

The formulations of REST and REMUCA are simple and straightforward, but
the numerical improvement is great, because the weight factor
determination for ST and MUCA becomes very difficult
by the usual iterative processes for complex systems.

\subsection{Simulated Tempering Replica-Exchange Method and
Multicanonical Replica-Exchange Method}

In the previous subsection we presented REST and REMUCA, 
which use a short REM run for the determination 
of the simulated temepering weight factor and 
the multicanonical weight factor, respectively. 
Here, we present two modifications of REM and refer to
the new methods as the 
{\it simulated tempering replica-exchange method} (STREM) \cite{STREM} 
and {\it multicanonical replica-exchange method} 
(MUCAREM) \cite{SO3,MSO03,MSO03b}.
In STREM
the production run is a REM simulation with a few replicas
that perform ST simulations with different temperature
ranges.
Likewise, in MUCAREM the production run is 
a REM simulation with a few replicas
in multicanonical ensembles, i.e.,
different replicas perform MUCA simulations with
different energy ranges.  
   
While ST and MUCA simulations are usually based on local
updates, a replica-exchange process can be considered to be
a global update, and global updates enhance the conformational
sampling further.

%----------------------
\section{MULTIDIMENSIONAL EXTENSIONS OF THE THREE GENERALIZED-ENSEMBLE
ALGORITHMS} 
%  \label{th:sec}
%----------------------
\subsection{General Formulations}
We now give the general formulations for the multidimensional
generalized-ensemble algorithms \cite{MO09a,MO09b,M09}. 
Let us consider a generalized potential energy function $E_{\bil}(x)$, which depends on 
$L$ parameters $\bil = (\lambda^{(1)}, \cdots, \lambda^{(L)})$, of a system in state $x$. 
Although $E_{\bil}(x)$ can be any function of $\bil$, we consider the following specific 
generalized potential energy function for simplicity:  
\begin{equation}
E_{\bil} (x) = 
E_0 (x) + \sum_{\ell = 1}^L \lambda^{(\ell)} V_{\ell} (x)~.
\label{EQN1}
\end{equation}
Here, there are $L+1$ energy terms, $E_0(x)$ and
$V_{\ell}(x)$ ($\ell =1, \cdots, L$), and
$\lambda^{(\ell)}$ are the
corresponding coupling constants for $V_{\ell}(x)$.  

After integrating out the momentum degrees of freedom, the partition function of the system at fixed temperature $T$ and parameters $\bil$ is given by
\begin{equation}
Z(T,\bil) = \int dx \exp(-\beta E_{\bil}(x))
=\dis \int dE_0 dV_1 \cdots dV_L ~ n(E_0,V_1,\cdots,V_L) 
\exp \left(-\beta E_{\bil} \right)~,
\label{EQN2}
\end{equation}
where $n(E_0,V_1,\cdots,V_L)$ is the multidimensional density of states: 
\begin{equation}
n(E_0,V_1,\cdots,V_L) 
=\dis \int dx \delta(E_0(x)-E_0) \delta(V_1(x)-V_1) \cdots \delta(V_L(x)-V_L)~. 
\label{eqn2.1}
\end{equation}
Here, the integration is replaced by a summation when $x$ is discrete. 

The expression in Eq.~(\ref{EQN1}) is 
often used in simulations. 
For instance, 
in simulations of spin systems, 
$E_0(x)$ and $V_{1}(x)$ (here, $L = 1$ 
and $x=\{ S_1, S_2, \cdots \}$ stand for spins) 
can be respectively considered as the zero-field term 
and the magnetization term coupled with
the external field $\lambda^{(1)}$. 
(For Ising model, $E_0 = -J \sum_{<i,j>} S_i S_j$,
$V_1 = - \sum_i S_i$, and $\lambda^{(1)}=h$, i.e., external magnetic field.) 
In umbrella sampling \cite{US} in molecular simulations, $E_0(x)$ and
$V_{\ell}(x)$ can be taken as the original potential
energy and the (biasing) umbrella potential energy, respectively, with the
coupling parameter $\lambda^{(\ell)}$ (here, $x= \{{\biq}_1, \cdots, {\biq}_N\}$ 
where ${\biq}_k$ is the coordinate
vector of the $k$-th particle and $N$ is the total number of particles).  
For the molecular simulations in the isobaric-isothermal ensemble,
$E_0(x)$ and $V_1(x)$ (here, $L = 1$) correspond respectively to the
potential energy $U$ and the volume ${\cal V}$ coupled with the pressure 
${\cal P}$. (Namely, we have
$x= \{{\biq}_1, \cdots, {\biq}_N, {\cal V}\}$, 
$E_0=U$, $V_1={\cal V}$, and 
$\lambda^{(1)}={\cal P}$, i.e., $E_{\bil}$ is the enthalpy without the kinetic energy contributions.)
For simulations in the grand canonical ensemble with $N$ particles,
we have $x= \{{\biq}_1, \cdots, {\biq}_N, N \}$, and 
$E_0(x)$ and $V_1(x)$ (here, $L = 1$) correspond respectively to the
potential energy $U$ and the total number of particles $N$ coupled with the 
chemical potential $\mu$. (Namely, we have
$E_0=U$, $V_1=N$, and
$\lambda^{(1)}=-\mu$.)

Moreover, going beyond the well-known ensembles discussed above, we can introduce any physical quantity of interest (or its function) as the additional potential energy term $V_{\ell}$. For instance, $V_{\ell}$ can be an overlap with a reference configuration in spin glass systems, an end-to-end distance, a radius of gyration in molecular systems, etc. 
In such a case, we have to carefully choose the range of $\lambda^{(\ell)}$ 
values so that the new energy term $\lambda^{(\ell)} V_{\ell}$ 
will have roughly the same order of magnitude as the original energy term $E_0$. 
We want to perform a simulation where a random walk not only in the $E_0$ space but also in the $V_{\ell}$ space is realized. As shown below, this can be done by performing a multidimensional REM, ST, 
or MUCA simulation. 

%\subsection{Multi-dimensional replica-exchange method}
We first describe the {\it multidimensional replica-exchange method} 
(MREM) \cite{SKO}. 
The crucial observation that led to this algorithm is:  
As long as we have $M$ {\it non-interacting}
replicas of the original system, the Hamiltonian 
$H(q,p)$ of the system does not have to be identical
among the replicas and it can depend on a parameter
with different parameter values for different replicas.
%%%%
The system for the multidimensional REM consists of $M$ non-interacting replicas of the original system in the
``canonical ensemble'' with $M (=M_0 \times M_1 \times \cdots \times M_L)$ different parameter 
sets $\biL_m$ ($m=1,\cdots,M$), 
where 
$\biL_{m} \equiv (T_{m_0},\bil_{m}) \equiv (T_{m_0},\lambda^{(1)}_{m_1},
\cdots,\lambda^{(L)}_{m_L})$ with $m_0=1, \cdots, M_0, m_{\ell}=1, 
\cdots, M_{\ell}$ ($\ell = 1, \cdots, L$).  
Because the replicas are non-interacting, the weight factor 
is given by the product of Boltzmann-like factors for each replica:
\begin{equation}
W_{\rm MREM}
\equiv \dis{\prod_{m_0=1}^{M_0} \prod_{m_1=1}^{M_1} \cdots \prod_{m_L=1}^{M_L}
 \exp \left(- \beta_{m_0} E_{\bil_m} \right)}~.
\label{eqn16}
\end{equation}

Without loss of generality we can order the parameters
so that $T_1 < T_2 < \cdots < T_{M_0}$ and
$\lambda^{(\ell)}_{1} < \lambda^{(\ell)}_{2} < \cdots < \lambda^{(\ell)}_{M_{\ell}}~$ (for each $\ell = 1,\cdots, L$).  
The multidimensional REM is realized by alternately performing the following two steps: 
\begin{enumerate}
\item  For each replica, a ``canonical'' MC or MD simulation at the fixed
parameter set is carried out simultaneously and independently for a certain steps. 
\item 
We exchange a pair of replicas $i$ and $j$ which are at the
parameter sets $\biL_m$ and $\biL_{m+1}$, respectively.
The transition probability for this replica exchange process is given by 
\begin{equation}
w(\biL_m \leftrightarrow \biL_{m+1})
={\rm min}\left(1,\exp(-\Delta)\right)~,
\label{eqn18}
\end{equation}
where we have
\begin{equation}
\Delta = \left(\beta_{m_0} - \beta_{m_0+1} \right) 
\left(E_{\bil_m} \left(q^{[j]}\right) - E_{\bil_m} \left(q^{[i]}\right) \right)~,
\label{Eqn31}
\end{equation}
for $T$-exchange, and 
\begin{equation}
\Delta = \beta_{m_0} \left[  \left( E_{\bil_{m_{\ell}+1}} (q^{[j]}) - E_{\bil_{m_{\ell}+1}} (q^{[i]}) \right)
- \left( E_{\bil_{m_{\ell}}} (q^{[j]}) - E_{\bil_{m_{\ell}}} (q^{[i]}) \right) \right]~,
\label{eqn32}
\end{equation}
for $\lambda^{(\ell)}$-exchange (for one of $\ell=1,\cdots,L$).
Here, $q^{[i]}$ and $q^{[j]}$ stand for configuration variables
for replicas $i$ and $j$, respectively, before the replica exchange.

\end{enumerate}

%\subsection{Multi-dimensional simulated tempering}
We now consider the {\it multidimensional simulated tempering} (MST)
which realizes a random walk both in temperature $T$ 
and in parameters $\bil$ \cite{MO09a,MO09b,M09}. 
The entire parameter set $\biL=(T,\bil) \equiv (T, \lambda^{(1)}, \cdots, \lambda^{(L)})$ 
become dynamical variables and both the configuration and the parameter set 
are updated during the simulation with a weight factor:
\begin{equation}
W_{\rm MST} (\biL)
\equiv \exp \left(-\beta E_{\bil} + f(\biL) \right)~,
\label{Eqn8}
\end{equation}
where the function $f(\biL)=f(T,\bil)$ is chosen so that 
the probability distribution of $\biL$ is flat:
\begin{equation}
P_{\rm MST}(\biL)\propto
\dis \int dE_0 dV_1 \cdots dV_L~ n(E_0,V_1,\cdots,V_L)~ 
\exp \left(-\beta E_{\bil} + f(\biL) \right) 
\equiv {\rm constant}~.
\lab{EQN9}
\end{equation}
This means that $f(\biL)$ is the dimensionless (``Helmholtz'') free energy: 
\begin{equation}
 \exp \left( -f(\biL) \right) 
= \dis \int dE_0 dV_1 \cdots dV_L ~ n(E_0,V_1,\cdots,V_L)~ \exp (-\beta E_{\bil})~. 
\lab{eqn10}
\end{equation}

In the numerical work we discretize the parameter
set $\biL$ in $M (= M_0 \times M_1  \times \cdots \times M_L)$ different values: 
$\biL_{m} \equiv (T_{m_0},\bil_{m}) \equiv (T_{m_0},\lambda^{(1)}_{m_1},\cdots,\lambda^{(L)}_{m_L})$, where $m_0=1, \cdots, M_0, m_{\ell}=1, \cdots, M_{\ell}$ ($\ell = 1, \cdots, L$).  
Without loss of generality we can order the parameters
so that $T_1 < T_2 < \cdots < T_{M_0}$ and
$\lambda^{(\ell)}_{1} < \lambda^{(\ell)}_{2} < \cdots < \lambda^{(\ell)}_{M_{\ell}}~$ (for each $\ell = 1,\cdots, L$).  
The free energy $f(\biL_m)$ is now written as $f_{m_0,m_1,\cdots,m_L}=f(T_{m_0},\lambda^{(1)}_{m_1},\cdots, \lambda^{(L)}_{m_L}$).

Once the initial configuration and
the initial parameter set are chosen,
the multidimensional ST is realized by alternately 
performing the following two steps:
\begin{enumerate}
\item A ``canonical'' MC or MD simulation at the fixed parameter
set $\biL_m = (T_{m_0},\bil_{m}) =$ \\
$(T_{m_0},\lambda^{(1)}_{m_1},\cdots, \lambda^{(L)}_{m_L})$ is carried 
out for a certain steps with the weight factor 
$\exp(-\beta_{m_0} E_{\bil_m})$ (for fixed $\biL_m$, $f(\biL_m)$ 
in Eq.~(\ref{Eqn8}) is a constant and does not contribute).
\item 
We update the parameter set $\biL_m$ to a new parameter set $\biL_{m \pm 1}$ in which one of the parameters in $\biL_{m}$ is changed to a neighboring value with the configuration and the other parameters fixed. 
The transition probability of
this parameter updating
process is given by the following Metropolis criterion:
\begin{equation}
w(\biL_{m} \rightarrow \biL_{m \pm 1})
= {\rm min}\left(1,\dis \frac{W_{\rm MST}(\biL_{m \pm 1})}
{W_{\rm MST}(\biL_{m})} \right)
= {\rm min}\left(1,\exp \left( - \Delta \right)\right)~.
\label{eqn11}
\end{equation}
Here, there are two possibilities for $\biL_{m \pm 1}$, namely, 
$T$-update and $\lambda^{(\ell)}$-update. 
For $T$-update, we have 
$\biL_{m \pm 1} = (T_{m_0 \pm 1}, \bil_m)$ 
with 
\begin{equation}
\Delta = \left(\beta_{m_0 \pm 1} - \beta_{m_0} \right) E_{ \bil_{ m } }
- \left(f_{m_0 \pm 1,m_1,\cdots,m_L} - f_{m_0,m_1,\cdots,m_L} \right)~. 
\label{eqn12}
\end{equation}
For $\lambda^{(\ell)}$-update (for one of $\ell=1,\cdots,L$), we have 
$\biL_{m \pm 1} = (T_{m_0}, \bil_{m_{\ell} \pm 1})$ with 
\begin{equation}
\begin{array}{rl}
\Delta = \beta_{m_0} (E_{\bil_{m_{\ell} \pm 1}} - E_{\bil_{m_{\ell}}}) 
- \left(f_{m_0,\cdots,m_{\ell} \pm 1,\cdots} - f_{m_0,\cdots,m_{\ell},\cdots} \right)~,
\end{array}
\label{eqn13a}
\end{equation}
where $\bil_{m_{\ell} \pm 1} = (\cdots,\lambda^{(\ell-1)}_{m_{\ell-1}},\lambda^{(\ell)}_{m_{\ell} \pm 1},\lambda^{(\ell+1)}_{m_{\ell+1}},\cdots)$ and 
$\bil_{m_{\ell}} = (\cdots,\lambda^{(\ell-1)}_{m_{\ell-1}},\lambda^{(\ell)}_{m_{\ell}},\lambda^{(\ell+1)}_{m_{\ell+1}},\cdots)$.  
\end{enumerate}

%%\subsection{Multi-dimensional multicanonical algorithm}
We now describe the {\it multidimensional multicanonical algorithm} (MMUCA)
which realizes 
a random walk in the $( L+1 )$-dimensional space of $E_0(x)$ and $V_{\ell}(x)$ ($\ell~=~1, \cdots, L$). 

In the multidimensional MUCA ensemble, 
each state is weighted by the MUCA weight factor 
$W_{\rm MMUCA}(E_0,V_1,\cdots, V_L)$ so that a uniform energy distribution 
of $E_0$, $V_1, \cdots$, and $V_L$ may be obtained:
\begin{equation}
P_{\rm MMUCA}(E_0,V_1,\cdots,V_L) 
\propto n(E_0,V_1,\cdots,V_L) W_{\rm MMUCA}(E_0,V_1,\cdots,V_L) 
\equiv {\rm constant}~,
\label{EQN3}
\end{equation}
where $n(E_0,V_1,\cdots,V_L)$ is the multidimensional density of states.
From this equation, we obtain
\begin{equation}
W_{\rm MMUCA}(E_0,V_1,\cdots,V_L) \equiv 
\exp \left(-\beta_a E_{\rm MMUCA}(E_0,V_1,\cdots,V_L) \right) 
\propto \frac{1}{n(E_0,V_1,\cdots,V_L)}~,
\label{EQN4}
\end{equation}
where we have introduced an arbitrary reference temperature, 
$T_a = 1/k_{\rm B} \beta_a$, and wrote the weight factor 
in the Boltzmann-like form. 
Here, the ``{\it multicanonical potential energy}'' is defined by
\begin{equation}
E_{\rm MMUCA}(E_0,V_1,\cdots,V_L;T_a)  
\equiv k_{\rm B} T_a \ln n(E_0,V_1,\cdots, V_L)~.
\label{EQN5}
\end{equation}
The multidimensional MUCA MC simulation can be performed with the
following Metropolis transition probability from state $x$ with 
energy $E_{\bil}=E_0 + \sum_{\ell=1}^{L} \lambda^{(\ell)} V_{\ell}$ to state $x^{\prime}$ with
energy ${E_{\bil}}^{\prime}={E_0}^{\prime} + \sum_{\ell=1}^{L} \lambda^{(\ell)} {V_{\ell}}^{\prime}$ :
\begin{equation}
 w(x \rightarrow x^{\prime})
= {\rm min} \left(1,\dis \frac{W_{\rm MMUCA}({E_0}^{\prime},{V_{1}}^{\prime},\cdots,{V_{L}}^{\prime})}
{W_{\rm MMUCA}(E_0,V_1,\cdots,V_{L})}\right) 
= {\rm min} \left(1,\dis \frac{n(E_0,V_1,\cdots,V_L)}{n({E_0}^{\prime},{V_{1}}^{\prime},\cdots,{V_{L}}^{\prime})} \right)~.
\label{EQN6}
\end{equation}
A MD algorithm in the multidimensional MUCA ensemble
also naturally follows from Eq.~(\ref{EQN4}), in which a
regular constant temperature MD simulation
(with $T=T_a$) is performed by replacing the total potential energy 
$E_{\bil}$ by the multicanonical potential energy $E_{\rm MMUCA}$ in 
Eq.~(\ref{eqn4f2}):
\begin{equation}
\dot{\bs{p}}_k ~=~ 
- \frac{\partial E_{\rm MMUCA}( E_0,V_1,\cdots,V_L;T_a)}{\partial \bs{q}_k}
- \frac{\dot{s}}{s}~\bs{p}_k~.
\label{EQN7}
\end{equation}

We remark that the random walk in 
$E_{0}$ and in $V_{\ell}$ for the MUCA simulation
corresponds to that in 
$\beta$ and in $\beta \lambda^{(\ell)}$ for the ST simulation:
\begin{equation}
\left\{
\begin{array}{ll}
E_0 & \longleftrightarrow \beta~, \cr
V_{\ell} & \longleftrightarrow \beta \lambda^{(\ell)}~,~~(\ell=1,\cdots,L)~.
\end{array}
\right.
\end{equation}
They are in conjugate relation. 
   
\subsection{Weight Factor Determinations for Multidimensional ST and MUCA}
%%%%% How to determine the parameter
Among the three multidimensional generalized-ensemble algorithms 
described above, only MREM can be performed without much preparation 
because the weight factor for MREM is just a product of regular Boltzmann-like factors. 
On the other hand, we do not know the MST and MMUCA weight 
factors {\it a priori} and need to estimate them. 
As a simple method for these weight factor determinations,
we can generalize the REST and REMUCA presented in the previous subsections 
to multidimensions.

Suppose we have made a single run of a short MREM simulation 
with $M (= M_0 \times M_1  \times \cdots \times M_L)$ replicas that correspond
to $M$ different parameter sets
$\biL_m$ ($m=1, \cdots, M$).
Let $N_{m_0,m_1,\cdots,m_L}(E_0,V_1,\cdots,V_L)$ and $n_{m_0,m_1,\cdots,m_L}$
be respectively 
the ($L+1$)-dimensional potential-energy histogram and the total number of
samples obtained for the $m$-th parameter set 
$\biL_m=(T_{m_0},\lambda^{(1)}_{m_1},\cdots,\lambda^{(L)}_{m_L})$. 
The generalized WHAM equations are then given by
\begin{equation}
n(E_0,V_1,\cdots,V_L) 
= \frac{\dis{\sum_{m_0,m_1,\cdots,m_L} 
N_{m_0,m_1,\cdots,m_L}(E_0,V_1,\cdots,V_L)}} 
{\dis{\sum_{m_0,m_1,\cdots,m_L} 
n_{m_0,m_1,\cdots,m_L}~\exp \left(f_{m_0,m_1,\cdots,m_L}-\beta_{m_0} 
E_{\bil_m}\right)}}~,
\label{eqn20}
\end{equation}
and 
\begin{equation}
\exp (-f_{m_0,m_1,\cdots,m_L})
= \dis{\sum_{E_0,V_1,\cdots,V_L} n(E_0,V_1,\cdots,V_L) \exp \left(-\beta_{m_0} E_{\bil_m}\right)}~. 
\label{eqn21}
\end{equation}
The density of states
$n(E_0,V_1,\cdots,V_L)$ (which is inversely proportional to the MMUCA
weight factor) and the dimensionless free energy $f_{m_0,m_1,\cdots,m_L}$ 
(which is the MST parameter) are obtained by 
solving Eqs.~(\ref{eqn20}) and (\ref{eqn21}) self-consistently by iteration.

\subsection{Expectation Values of Physical Quantities}
We now present the equations to calculate ensemble averages of physical quantities with 
any temperature $T$ and any parameter $\bil$ values. 
   
After a long production run of MREM and MST simulations, 
the canonical expectation value of a physical quantity $A$
with the parameter values
$\biL_m$ ($m=1,\cdots,M$), 
where 
$\biL_{m} \equiv (T_{m_0},\bil_{m}) \equiv (T_{m_0},\lambda^{(1)}_{m_1},
\cdots,\lambda^{(L)}_{m_L})$ with $m_0=1, \cdots, M_0, m_{\ell}=1, 
\cdots, M_{\ell}$ ($\ell = 1, \cdots, L$), and
$M (=M_0 \times M_1 \times \cdots \times M_L)$, can be
calculated by the usual arithmetic mean:
\begin{equation}
<A>_{T_{m_0},\bil_{m}} = \frac{1}{n_{m}} \sum_{k=1}^{n_{m}}
A\left(x_{m}(k)\right)~,
\label{Eqn7b}
\end{equation}
where $x_m(k)$ ($k=1,\cdots,n_m$) are the configurations 
obtained with the parameter values
$\biL_m$ ($m=1,\cdots,M$), 
and $n_{m}$ is the total number of measurements made
with these parameter values.
The expectation values of $A$ at any 
intermediate $T$ ($=1/k_{\rm B} \beta$) and 
any $\bil$ can also be obtained from
\begin{equation}
<A>_{T,\bil} \ = \frac{ \dis \sum_{E_0,V_1,\cdots,V_L}
A(E_0,V_1,\cdots,V_L) n(E_0,V_1,\cdots,V_L) \exp \left(-\beta E_{\bil}\right) }
{\dis \sum_{E_0,V_1,\cdots,V_L} n(E_0,V_1,\cdots,V_L) 
\exp \left(-\beta E_{\bil}\right) }~,
\label{Eqn29}
\end{equation}
where the density of states $n(E_0,V_1,\cdots,V_L)$ is obtained
from the multiple-histogram reweighting techniques.
Namely, from the MREM or MST simulation, we first obtain 
the histogram $N_{m_0,m_1,\cdots,m_L}(E_0,V_1,\cdots,V_L)$ 
and the total number of samples $n_{m_0,m_1,\cdots,m_L}$ 
in Eq.~(\ref{eqn20}). 
The density of states
$n(E_0,V_1,\cdots,V_L)$ and the dimensionless 
free energy $f_{m_0,m_1,\cdots,m_L}$ are then 
obtained by solving Eqs.~(\ref{eqn20}) and (\ref{eqn21}) 
self-consistently by iteration.
Substituting the obtained density of states $n(E_0,V_1,\cdots,V_L)$ 
into Eq.~(\ref{Eqn29}), one can calculate the ensemble average of 
the physical quantity 
$A$ at any $T$ and any $\bil$. 

Moreover, the ensemble average of the physical quantity $A$ 
(including those that cannot be expressed 
as functions of $E_0$ and $V_{\ell}$ ($\ell=1, \cdots, L$) ) can be 
obtained from the 
``trajectory'' of configurations of the production run \cite{MO09b}. 
Namely, we first obtain $f_{m_0,m_1,\cdots,m_L}$ 
for each $(m_0=1,\cdots,M_0, m_1=1,\cdots,M_1, \cdots, m_L=1,\cdots,M_L)$ by 
solving Eqs.~(\ref{eqn20}) and (\ref{eqn21}) self-consistently, 
and then we have
\begin{equation}
<A>_{T,\bil} \ = \frac{ \dis \sum_{m_0=1}^{M_0} \cdots \sum_{m_L=1}^{M_L} 
\sum_{x_m} A(x_m) 
\frac{\dis \exp \left( -\beta E_{\bil}(x_m) \right)}
{ \dis \sum_{n_0=1}^{M_0} \cdots \sum_{n_L=1}^{M_L} 
n_{n_0,\cdots,n_L}~
\exp \left( f_{n_0,\cdots,n_L}- \beta_{n_0} E_{\bil_n}(x_m) \right) } }
{ \dis \sum_{m_0=1}^{M_0} \cdots \sum_{m_L=1}^{M_L} \sum_{x_m} 
\frac{\dis \exp \left( -\beta E_{\bil}(x_m) \right)}
{ \dis \sum_{n_0=1}^{M_0} \cdots \sum_{n_L=1}^{M_L} 
n_{n_0,\cdots,n_L}~
\exp \left( f_{n_0,\cdots,n_L}- \beta_{n_0} E_{\bil_n}(x_m) \right) } }~,
\label{eqn34}
\end{equation}
where $x_m$ are the configurations obtained at 
$\biL_{m} = (T_{m_0},\bil_{m}) = (T_{m_0},\lambda^{(1)}_{m_1},
\cdots,\lambda^{(L)}_{m_L})$.  Here, the trajectories
$x_m$ are stored for each
$\biL_{m}$ separately.
    
For the MMUCA simulation with the weight 
factor $W_{\rm MMUCA}(E_0,\cdots,V_L)$, 
the expectation values of $A$ at any 
$T$ ($=1/k_{\rm B} \beta$) and any $\bil$ can also be obtained from
Eq.~(\ref{Eqn29}) by the single-histogram reweighting techniques
as follows.
Let $N_{\rm MMUCA}(E_0,V_1,\cdots,V_L)$ be the histogram of the distribution 
of $E_0,V_1,\cdots,V_L$, $P_{\rm MMUCA}(E_0,V_1,\cdots,V_L)$, obtained by 
the production run. 
The best estimate of the density of states $n(E_0,V_1,\cdots,V_L)$ is then
given by
\begin{equation} 
n(E_0,V_1,\cdots,V_L)
= \displaystyle{\frac{
N_{\rm MMUCA}(E_0,V_1,\cdots,V_L)}{ W_{\rm MMUCA}(E_0,\cdots,V_L)}}~. 
\label{eqnSH}
\end{equation}
Moreover, the ensemble average of the physical quantity $A$ (including those that cannot be expressed as 
a function of $E_0$ and $V_{\ell}$ ($\ell=1,\cdots,L$)) can be obtained 
as long as one stores the 
``trajectory'' of configurations $x_k$ from the production run. 
We have 
\begin{equation}
<A>_{T,\bil}=\frac{\dis \sum_{k=1}^{n_s} A(x_k) 
W_{\rm MMUCA}^{-1}(E_0(x_k),\cdots,V_L(x_k))
\exp \left(-\beta E_{\bil}(x_k) \right) } 
{\dis \sum_{k=1}^{n_s} 
W_{\rm MMUCA}^{-1}(E_0(x_k),\cdots,V_L(x_k))
\exp \left(-\beta E_{\bil}(x_k) \right) }~. 
\end{equation}
Here, $x_k$ is the configuration at the $k$-th MC (or MD) step
and $n_s$ is the total number of configurations
stored.

\subsection{Multidimensional Generalized-Ensemble Algorithms for 
the Isobaric-Isothermal Ensemble}
As examples of the multidimensional formulations in the previous subsections, 
we present the generalized-ensemble algorithms for the
isobaric-isothermal ensemble (or, the NPT ensemble) \cite{MO-10b}.
Let us consider a physical system that consists of $N$ atoms and 
that is in a box of a finite volume $\cV$.
The states of the system are specified by 
coordinates $q\equiv \{\biq_1,\biq_2,\cdots, \biq_N\}$ 
and momenta $p\equiv \{\bip_1,\bip_2,\cdots,\bip_N\}$ 
of the atoms and volume $\cV$ of the box.
The potential energy $E(q,\cV)$ for the system is a function of $q$ and $\cV$.

In the isobaric-isothermal ensemble 
\cite{nose84,nosejcp84,Andersen,mcd72} 
the probability distribution $P_{\rm NPT}(E,{\cal V};T,{\cal P})$ 
for potential energy $E$ and volume ${\cal V}$ at
temperature $T$ and pressure ${\cal P}$ is given by
\begin{equation}
 P_{\rm NPT}(E,{\cal V};T,{\cal P}) \propto n(E,{\cal V}) W_{\rm NPT}(E,{\cal V};T,{\cal P}) 
=n(E,{\cal V}) {\rm e}^{-\beta {\cal H}}~. 
\label{mbt-eqn2}
\end{equation}
Here, the density of states $n(E,{\cal V})$ is given
as a function of both $E$ and ${\cal V}$, 
and ${\cal H}$ is the ``enthalpy'' (without the kinetic energy contributions):
\begin{equation}
 {\cal H} = E + {\cal P}{\cal V}~. 
\label{mbt-eqn3}
\end{equation}
This weight factor produces an isobaric-isothermal ensemble
at constant temperature ($T$) and constant pressure (${\cal P}$). 
Note that this is a special case of the general formulations
in Eq.~(\ref{EQN1}) with $L=1$, $E_0=E$, $V_1=\cV$, 
and $\lambda^{(1)}={\cal P}$.

In order to perform the isobaric-isothermal MC simulation~\cite{mcd72},
we perform Metropolis sampling on the scaled coordinates 
$\sigma= \{ \bisg_1,\bisg_2,\cdots,\bisg_N\}$ 
where $\bisg_k = \cV^{-1/3}\biq_k \ (k=1,2,\cdots,N)$
($\biq_k$ are the real coordinates) 
and the volume ${\cal V}$ (here, the particles are placed in a cubic box of
a side of size $\cV^{-1/3}$).
The trial moves from state $x$ with the scaled coordinates 
$\sigma$ with volume ${\cal V}$ to state $x^{\prime}$ with the scaled coordinate 
$\sigma'$ and 
volume ${\cal V}'$ are generated by uniform random numbers. 
The enthalpy is accordingly changed 
from ${\cal H}(E(\sigma,{\cal V}),{\cal V})$ 
to ${\cal H}'(E(\sigma',{\cal V}'),{\cal V}')$ 
by these trial moves. 
The trial moves will be accepted with the following Metropolis criterion:
\begin{equation}
w(x \rightarrow x^{\prime})
= {\rm min} \left(1,\exp[-\beta \{ 
 {\cal H}' - {\cal H} - N k_{\rm B} T \ln({\cal V}'/{\cal V})
 \}]\right)~, 
 \label{accibt:eq}
\end{equation}
where $N$ is the total number of atoms in the system.

As for the MD method in this ensemble,
we just present the Nos\'e-Andersen algorithm~\cite{nose84,nosejcp84,Andersen}.
The equations of motion in Eqs.~(\ref{eqn4f1})--(\ref{eqn4f})
are now generalized as follows:
\begin{eqnarray}
\dot{\bs{q}}_k &=& \frac{\bs{p}_k}{m_k}
+ \frac{\dot{{\cal V}}}{3{\cal V}}~\bs{q}_k~, \label{isobath-a} \\
\dot{\bs{p}}_k &=&
- \frac{\partial {\cal H}}{\partial \bs{q}_k}
- \left( \frac{\dot{s}}{s} + \frac{\dot{{\cal V}}}{3{\cal V}} \right)
\bs{p}_k 
\label{isobath-b} \\
&=& \bs{f}_k 
- \left( \frac{\dot{s}}{s} + \frac{\dot{{\cal V}}}{3{\cal V}} \right)
\bs{p}_k~, \label{isobath-b2} \\
\dot{s} &=& s~\frac{P_s}{Q}~, \\
\dot{P}_s &=&
\sum_{i=1}^{N} \frac{\bs{p}_i^{2}}{m_i} - 3N k_{\rm B} T
~=~
3Nk_{\rm B} \left( T(t) - T \right)~, \\
\dot{{\cal V}} &=& s~\frac{P_{\cV}}{\cM}~, \\
\dot{P}_{\cV} &=&
s \left[\frac{1}{3{\cal V}}
\left( \sum_{i=1}^{N} \frac{\bs{p}_i^{2}}{m_i}
- \sum_{i=1}^{N} \bs{q}_i \cdot
  \frac{\partial {\cal H}}{\partial \bs{q}_i}
  \right)
- \frac{\partial {\cal H}}{\partial {\cal V}} \right]
\label{isobath-c} \\
&=& s \left( {\cal P}(t) - {\cal P} \right)~, \label{isobath-c2}
\end{eqnarray}
where $\cM$ is the artificial mass associated with the volume,
$P_{\cV}$ is the conjugate momentum for the volume, and
the ``instantaneous pressure'' ${\cal P}(t)$ is defined by
\begin{equation}
{\cal P}(t) = 
\frac{1}{3{\cal V}}
\left( \sum_{i=1}^{N} \frac{\bs{p}_i(t)^{2}}{m_i}
+ \sum_{i=1}^{N} \bs{q}_i(t) \cdot \bs{f}_i(t)
  \right)  
- \frac{\partial E}{\partial {\cal V}}(t)~.
\end{equation}

In REM simulations for the NPT ensemble, we prepare a system that 
consists of $M_T \times M_{\cP}$ 
non-interacting replicas of the original system, where $M_T$ and $M_{\cP}$ are 
the number of temperature and pressure values used in the simulation, respectively. 
The replicas are specified by labels $i$ $(i=1,2,\cdots ,M_T\times M_{\cP})$, 
temperature by $m_0$ $(m_0=1,2,\cdots,M_T)$ and pressure by $m_1$ $(m_1=1,2,\cdots,M_{\cP})$.

To perform REM simulations, we carry out the following two steps 
alternately: (1) perform a usual constant NPT MC or MD simulation 
in each replica 
at assigned temperature and pressure and (2) try to exchange the replicas.
If the temperature (specified by $m_0$ and $n_0$) or pressure (specified 
by $m_1$ and $n_1$) between the replicas is exchanged in Step 2, 
the transition probability from 
$X\equiv \{\cdots,(\sigma^{[i]},\cV^{[i]};T_{m_0},\cP_{m_1}),
\cdots,(\sigma^{[j]},\cV^{[j]};T_{n_0},\cP_{n_1} ),\cdots\}$ 
to $X'\equiv \{\cdots,(\sigma^{[i]},\cV^{[i]};T_{n_0},\cP_{n_1}),\cdots,$
$(\sigma^{[j]},\cV^{[j]};T_{m_0},\cP_{m_1} ),\cdots\}$ 
at the trial is given by~\cite{OKOM,RevSO}
\begin{equation}
w_{\rm{REM}}(X \to X') = \min \left[ 1, \exp (-\Delta _{\rm{REM}} ) \right] ,
	\label{rem_transition1}
\end{equation}
where
\begin{equation}
\Delta _{\rm{REM}} = (\beta_{m_0} - \beta_{n_0})
\left[ E(\sigma^{[j]},\cV^{[j]}) - E(\sigma^{[i]},\cV^{[i]}) \right] 
	+ (\beta_{m_0} \cP_{m_1}- \beta_{n_0} \cP_{n_1})\left( \cV^{[j]} - \cV^{[i]} \right) .
	\label{rem_transition2}
\end{equation}

In ST simulations for the NPT ensemble, we introduce a function 
$f(T,\cP)$ and use a weight 
factor $W_{\rm{ST}}(E,\cV;T,\cP) \equiv \exp[-\beta (E+\cP \cV) + f(T,\cP)]$ 
so that the distribution function $P_{\rm{ST}}(T,\cP)$ of $T$ and $\cP$ may be uniform:
\begin{equation}
	P_{\rm{ST}}(T,\cP) \propto \int_0^{\infty}d\cV\int_{\cV} dq \ 
W_{\rm{ST}}[E(q,\cV),\cV;T,\cP] = \rm{constant}~.
	\label{eq:dis:st}
\end{equation}
From Eq.~(\ref{eq:dis:st}), it is found that $f(T,\cP)$ is formally given by
\begin{equation}
	f(T,\cP) = -\ln \left\{ \int_0^{\infty}d\cV\int_{\cV}dq \exp \left[ -\beta \left( E(q,\cV) + 
\cP \cV \right) \right] \right\},
\end{equation}
and this function is the dimensionless Gibbs free energy except for a constant. 

To perform ST simulations, we again discretize temperature and pressure
into $M_0 \times M_1$ set of values $(T_{m_0},\cP_{m_1})$ 
$(m_0=1, \cdots, M_0, ~m_1=1, \cdots, M_1)$.
We carry out the following two steps 
alternately: (1) perform a usual constant NPT MC or MD simulation and 
(2) try to update the temperature or pressure. 
In Step 2 the transition probability from 
the state $X\equiv \{ \sigma,\cV;T_{m_0},\cP_{m_1}\}$ to 
the state $X'\equiv \{ \sigma,\cV;T_{n_0},\cP_{n_1} \}$ 
is given by 
\begin{equation}
w_{\rm{ST}}(X \to X') = \min \left[ 1, \exp (-\Delta _{\rm{ST}} ) \right],
	\label{st_transition1}
\end{equation}
where
\begin{equation}
\Delta _{\rm{ST}} = (\beta_{n_0} - \beta_{m_0})E(\sigma,\cV) + 
(\beta_{n_0} \cP_{n_1} - \beta_{m_0} \cP_{m_1})\cV - 
\left( f_{n_0,n_1} - f_{m_0,m_1} \right)~.
	\label{st_transition2}
\end{equation}

We remark that when we perform MD simulations with REM and ST, 
the momenta should be 
rescaled if the replicas are exchanged for the temperature 
in REM and the temperature is updated in ST as shown above in the
previous subsections.
    
From the production run of REM or ST simulations in the NPT ensemble,
we can calculate isobaric-isothermal averages of a physical quantity
$A$ at $(T_{m_0},\cP_{m_1})$ $(m_0=1, \cdots, M_0, ~m_1=1, \cdots, M_1)$
by the usual arithmetic mean:
\begin{equation}
<A>_{T_{m_0},\cP_{m_1}} = \frac{1}{n_{m}} \sum_{k=1}^{n_{m}}
A\left(x_{m}(k)\right)~,
\label{Eqn7p}
\end{equation}
where $x_m(k)$ ($k=1,\cdots,n_m$) are the configurations 
obtained with the parameter values
$(T_{m_0},\cP_{m_1})$ 
and $n_{m}$ is the total number of measurements made
with these parameter values.
The expectation values of $A$ at any 
intermediate temperature $T$ ($=1/k_{\rm B} \beta$) and 
any intermediate pressure $\cP$ can also be obtained from
\begin{equation}
<A>_{T,\cP} \ = \frac{ \dis \sum_{E,\cV}
A(E,\cV) n(E,\cV) \exp \left(-\beta (E + \cP \cV) \right) }
{\dis \sum_{E,\cV} 
n(E,\cV) \exp \left(-\beta (E + \cP \cV) \right) }~,
\label{Eqn29p}
\end{equation}
where the density of states $n(E,\cV)$ is obtained
from the multiple-histogram reweighting techniques.
Namely, from the REM or ST simulation, we first obtain 
the histogram $N_{m_0,m_1}(E,\cV)$ 
and the total number of samples $n_{m_0,m_1}$. 
The density of states
$n(E,\cV)$ and the dimensionless 
free energy $f_{m_0,m_1}$ are then 
obtained by solving the following equations
self-consistently by iteration
(see Eqs.~(\ref{eqn20}) and (\ref{eqn21}) above):
\begin{equation}
n(E,\cV) 
= \frac{\dis{\sum_{m_0,m_1} 
N_{m_0,m_1}(E,\cV)}} {\dis{\sum_{m_0,m_1} 
n_{m_0,m_1}~\exp \left(f_{m_0,m_1}-\beta_{m_0} 
(E + \cP_{m_1} \cV) \right)} }~,
\label{eqn20p}
\end{equation}
and 
\begin{equation}
\exp (-f_{m_0,m_1})
= \dis{\sum_{E,\cV} n(E,\cV) \exp \left(-\beta_{m_0} 
(E + \cP_{m_1} \cV) \right) }~.
\label{eqn21p}
\end{equation}
Substituting the obtained density of states $n(E,\cV)$ 
into Eq.~(\ref{Eqn29p}), one can calculate the ensemble average of 
the physical quantity 
$A$ at any $T$ and any $\cP$. 

We now introduce the multicanonical algorithm
into the isobaric-isothermal ensemble and 
refer to this generalized-ensemble algorithm as 
the {\it multibaric-multithermal algorithm} (MUBATH)
\cite{OO03}--\cite{okokmd06}. 
The molecular simulations in this generalized ensemble perform random walks 
both in the potential energy space and in the volume space.
   
In the MUBATH ensemble, 
each state is sampled by the MUBATH weight factor 
$W_{\rm mbt}(E,{\cal V})\equiv \exp \{-\beta_a {\cal H}_{\rm mbt}(E,{\cal V})\}$ 
(${\cal H}_{\rm mbt}$ is referred to as the multibaric-multithermal enthalpy) 
so that a uniform distribution in both potential energy $E$
and volume ${\cal V}$
is obtained \cite{OO03}: 
\begin{equation}
 P_{\rm mbt}(E,{\cal V}) \propto n(E,{\cal V}) W_{\rm mbt}(E,{\cal V}) 
= n(E,{\cal V}) \exp \{-\beta_a {\cal H}_{\rm mbt}(E,{\cal V})\} 
\equiv {\rm constant}~, 
\label{mbt-eqn5}
\end{equation}
where we have chosen an arbitrary reference
temperature, $T_a = 1/k_{\rm B} \beta_a$. 

The MUBATH MC simulation can be performed by 
replacing ${\cal H}$ by ${\cal H}_{\rm mbt}$ in Eq.~(\ref{accibt:eq}):
\begin{equation}
w(x \rightarrow x^{\prime})
= {\rm min} \left(1,\exp[-\beta_a \{ 
 {\cal H}'_{\rm mbt} - {\cal H}_{\rm mbt} - N k_{\rm B} T_a \ln({\cal V}'/{\cal V})
 \}]\right)~, 
 \label{accmbt:eq}
\end{equation}

In order to perform the MUBATH MD simulation, 
we just solve the above equations of motion 
(Eqs.~(\ref{isobath-a})--(\ref{isobath-c2}))
for the regular isobaric-isothermal ensemble (with arbitrary reference
temperature $T=T_a$), where
the enthalpy ${\cal H}$ is replaced by the multibaric-multithermal enthalpy
${\cal H}_{\rm mbt}$ in 
Eqs.~(\ref{isobath-b}) and (\ref{isobath-c})~\cite{okokmd04}. 

In order to calculate the isobaric-isothermal-ensemble averages,
we employ the single-histogram reweighting techniques \cite{FS1}.
The expectation value of a physical quantity $A$
at any $T$ and any ${\cal P}$ is obtained 
by substituting the following density of states into Eq.~(\ref{Eqn29p}):
\begin{equation}
 n(E,{\cal V}) =
 \frac{N_{\rm mbt}(E,{\cal V})}{W_{\rm mbt}(E,{\cal V})}~,
\label{mbtrwt:eq}
\end{equation}
where $N_{\rm mbt}(E,{\cal V})$ is the histogram of 
the probability distribution
$P_{\rm mbt}(E,{\cal V})$
of potential energy and volume
that was obtained by the MUBATH production run.

%
      
%

%\section{COMPUTATIONAL  DETAILS}    
%\noindent
%{\bf RESULTS} \\
\section{EXAMPLES OF SIMULATION RESULTS}
We tested the effectiveness of the generalized-ensemble
algorithms by using a system of
a 17-residue fragment of ribonuclease $T_1$ \cite{SCH,MSO03b}.  
It is known by experiments that this peptide fragment
forms $\alpha$-helical conformations \cite{SCH}.  
We have performed a two-dimensional REM simulation and a two-dimensional ST simulation.
In these simulations, we used the following energy function: 
\begin{equation}
E_{\lambda} = E_0 + \lambda E_{\rm SOL}~,
\label{Eqn16p}
\end{equation}
where we set $L=1, V_1=E_{\rm SOL}$, and $\lambda^{(1)}=\lambda$ 
in Eq.~(\ref{EQN1}). 
Here, $E_0$ is the potential energy of the solute 
and $E_{\rm SOL}$ is the solvation free energy. 
The parameters in the conformational energy 
as well as the molecular geometry were taken from ECEPP/2 \cite{ECEPP1,ECEPP2,ECEPP3}. 

The solvation term $E_{\rm SOL}$ is given by the sum of terms that are proportional to the solvent-accessible surface area of 
heavy atoms of the solute \cite{SOL1}.
For the calculations of solvent-accessible surface area, we used the 
computer code NSOL \cite{SOL2}. 

The computer code KONF90 \cite{KONF1,KONF2} was modified in order to accommodate the generalized-ensemble algorithms. 
The simulations were started from randomly generated conformations. 
We prepared eight temperatures ($M_0=$8) which are distributed exponentially between $T_1=$ 300 K and $T_{M_0}=$ 700 K (i.e., 300.00, 338.60, 382.17, 431.36, 486.85, 549.49, 620.20, and 700.00 K) and four equally-spaced $\lambda$ values ($M_1=4$) ranging from 0 to 1 (i.e., $\lambda_1$ = 0, $\lambda_2$ = 1/3, $\lambda_3$ = 2/3, and $\lambda_4$ = 1) in the two-dimensional REM simulation and the two-dimensional ST simulation.
Simulations with $\lambda$ = 0 (i.e., $E_{\lambda} = E_0$ ) and with 
$\lambda$ = 1 (i.e., $E_{\lambda} = E_0 + E_{\rm SOL}$) correspond to those in gas phase and in aqueous solution, respectively. 

We first present the results of the two-dimensional REM simulation. We used 32 replicas with 
the eight temperature values and the four $\lambda$ values given above. 
Before taking the data, we made the two-dimensional REM simulation of 100000 MC sweeps with each replica for thermalization. 
We then performed the two-dimensional REM simulation of 1000000 MC sweeps for each replica to determine the weight factor for the two-dimensional ST simulation. 
At every 20 MC sweeps, either $T$-exchange or $\lambda$-exchange was tried 
(the choice of $T$ or $\lambda$ was made randomly). 
In each case, either set of pairs of replicas ((1,2),...,($M-$1,$M$)) or ((2,3),...,($M$,1)) was also chosen randomly, 
where $M$ is $M_0$ and $M_1$ for $T$-exchange and $\lambda$-exchange, respectively. 

%\begin{figure}[htd]
\begin{figure}[htbp]
\begin{center}
\includegraphics[width=11.0cm]{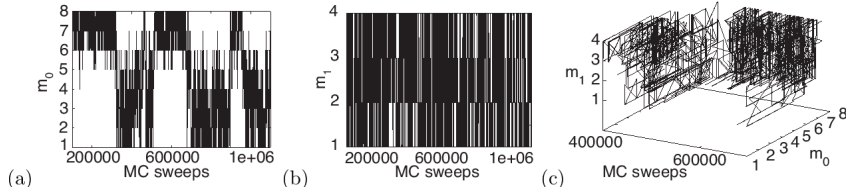}
%\vspace{2cm}
\caption{Time series of the labels of 
$T_{m_0}$, $m_0$, (a) and 
$\lambda_{m_1}$, $m_1$, (b) as functions of MC sweeps, 
and that of both $m_0$ and $m_1$ for the 
region from 400000 MC sweeps to 700000 MC sweeps (c). 
The results were from one of the replicas (Replica 1). 
In (a) and (b), MC sweeps start at 100000 and end 
at 1100000 because the first 100000 sweeps have been removed 
from the consideration for thermalization purpose.} 
\end{center}
\label{Fig1}
\end{figure}

%In Fig.~\ref{Fig1} we show the time series of labels 
In Fig.~1 we show the time series of labels 
of $T_{m_0}$ (i.e., $m_0$) and $\lambda_{m_1}$ (i.e., $m_1$) for one of the replicas. 
The replica realized a random walk not only in temperature space but also in $\lambda$ space. 
The behavior of $T$ and $\lambda$ for other replicas was also similar 
%(data not shown). 
(see Ref.~\cite{M09}). 
From Fig.~1, one finds that the $\lambda$-random walk is more frequent 
than the $T$-random walk. 

%\begin{figure}[th]
\begin{figure}[htbp]
\begin{center}
\includegraphics[width=11.0cm]{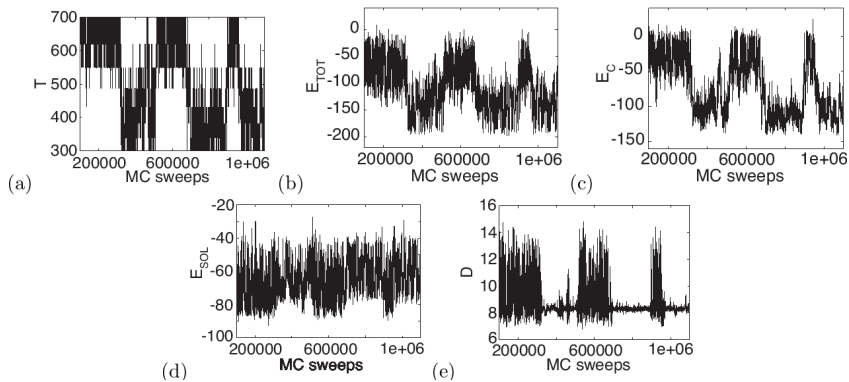}
\caption{Time series of the temperature 
$T$ (a), total energy 
$E_{\rm TOT}$ (b), conformational energy $E_{\rm C}$ (c), 
solvation free energy $E_{\rm SOL}$ (d), 
and end-to-end distance $D$ (e) for the same replica as 
%in Fig.~\ref{Fig1}.
in Fig.~1.
The temperature is in K, the energy is in kcal/mol, 
and the end-to-end distance is in \AA.}
\end{center}
\label{Fig2}
\end{figure}

We also show the time series of temperature $T$, total energy $E_{\rm TOT}$, 
conformational energy $E_{\rm C}$, solvation free energy $E_{\rm SOL}$, 
and end-to-end distance $D$ for the same replica in Fig.~2. 
From Figs.~2(a) and 2(e), we find that at lower temperatures 
the end-to-end distance is about 8 \AA, which is the length of a 
fully $\alpha$-helical conformation and that at higher temperatures it fluctuates 
much for a range from 7 \AA \ to 14 \AA. 
It suggests that $\alpha$-helix structures exist at low temperatures and 
random-coil structures occur at high temperatures. 
There are transitions from/to $\alpha$-helix structures to/from random coils during the simulation.
It indicates that the REM simulation avoided getting trapped in local-minimum-energy 
states and sampled a wide conformational space. 

%\begin{figure}[th]
\begin{figure}[htbp]
\begin{center}
\includegraphics[width=9.0cm]{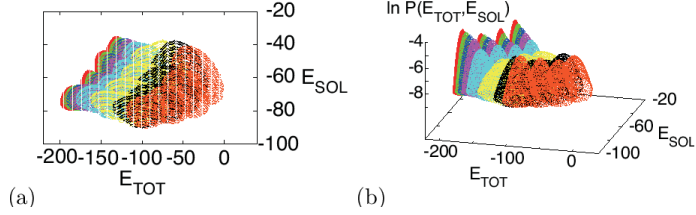}
\caption{Contour curves and histograms 
of distributions of the 
total energy $E_{\rm TOT}$ and the solvation free energy 
$E_{\rm SOL}$ ((a) and (b)) from the two-dimensional 
REM simulation.}
\end{center}
\label{Fig3}
\end{figure}
    
The canonical probability distributions of $E_{\rm TOT}$ and $E_{\rm SOL}$ at 
the 32 conditions obtained from the two-dimensional REM simulation 
%are shown in Fig.~\ref{Fig3}. 
are shown in Fig.~3. 
For an optimal performance of the REM simulation, there should be enough 
overlaps between all pairs of neighboring distributions, which will lead to 
sufficiently uniform and large acceptance ratios of replica exchanges. 
%There are indeed ample overlaps between the neighboring distributions in Fig.~\ref{Fig3}. 
There are indeed ample overlaps between the neighboring distributions in Fig.~3.

%\subsection{Two-Dimensional ST Simulation}
We now use the results of the two-dimensional REM simulation to determine the weight factors for the two-dimensional ST simulation by the multiple-histogram reweighting techniques. 
Namely, by solving the generalized WHAM equations 
in Eqs.~(\ref{eqn20}) and (\ref{eqn21}) with the obtained histograms at the 32 conditions 
%(see Fig.~\ref{Fig3}), 
(see Fig.~3), 
we obtained 32 values of the ST parameters $f_{m_0,m_1} (m_0=1,\cdots,8; m_1=1,\cdots,4)$. 

After obtaining the ST weight factor, 
$W_{ST}=\exp(-\beta_{m_0}(E_{\rm C}+\lambda_{m_1} E_{\rm SOL}) + f_{m_0,m_1})$, 
we carried out the two-dimensional ST simulation of 1000000 MC sweeps for data 
collection after 100000 MC sweeps for thermalization. 
At every 20 MC sweeps, either $T_{m_0}$ or $\lambda_{m_1}$ was respectively 
updated to $T_{m_0 \pm 1}$ or $\lambda_{m_1 \pm 1}$ (the choice 
of $T$ or $\lambda$ update and the choice of $\pm 1$ were made randomly). 

We show the average total energy, average conformational energy, 
average $\lambda \times E_{\rm SOL}$, and average end-to-end distance in Fig.~4. 
The results are in good agreement with those of the REM simulation (data not shown).
%in Fig.~\ref{Fig4}. 
%Note that we obtained similar results with only one replica (the CPU time were reduced to 1/32. 

%\begin{figure}[th]
\begin{figure}[htbp]
\begin{center}
\includegraphics[width=9.0cm]{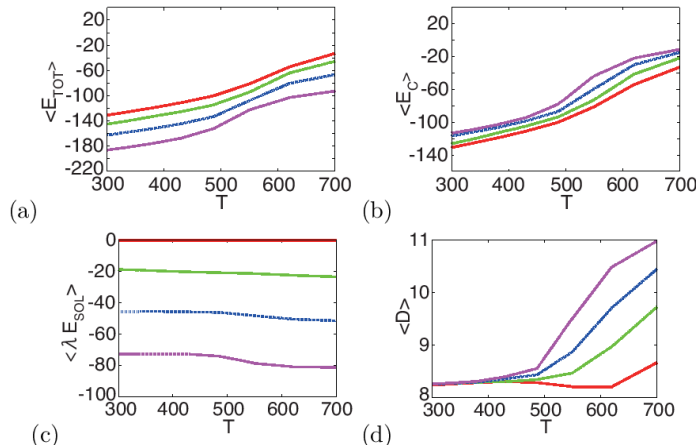}
\caption{The average total energy (a), 
average conformational energy (b), 
average of $\lambda \times E_{\rm SOL}$ (c), and average end-to-end 
distance (d) with all the $\lambda$ values as functions of temperature. 
The lines colored in red, green, blue, and purple are for $\lambda_1$, 
$\lambda_2$, $\lambda_3$, and $\lambda_4$, respectively.
They are in order from above to below in (a) and (c)
and from below to above in (b) and (d).}
\end{center}
\label{Fig9}
\end{figure}
    
We found that the results of the two-dimensional ST simulation are 
in complete agreement with those of the two-dimensional REM simulation 
for the average quantities.
The only difference between the two simulations is the number of replicas. 
In the present simulation, while the REM simulation 
used 32 replicas, 
the ST simulation 
used only one replica. 
Hence, we can save much computer power with ST. 
     
A second example of our multidimensional generalized-ensemble
simulations is a pressure ST (PST) simulation in the
isobaric-isothermal ensemble \cite{MO-10b}.
This simulation performs a random walk in one-dimensional pressure space.
The system that we simulated is ubiquitin in explicit water.
This system has been studied by high pressure NMR experiments and
known to undergo high-pressure denaturations \cite{KA03,KYA05}.
Ubiquitin has 76 amino acids and it was placed in a cubic box
of 6232 water molecules.
Temperature was fixed to be 300 K throughout the simulations, and
we prepared 100 values of pressure ranging from 1 bar to 10000 bar.
Temperature and pressure were controlled by Hoover-Langevin method \cite{HL}
and particle mesh Ewald method \cite{PME1,PME2} were employed 
for electrostatic interactions.
The time step was 2.0 fsec.
The force field CHARMM22 \cite{CHARMM22} with CMAP \cite{CMap1,CMap2} 
and TIP3P water model \cite{TIP3P,CHARMM22} were used, and 
the program package NAMD version 2.7b3 \cite{NAMD}
was modified to incorporate the PST algorithm.

We first performed 100 independent conventional isobaric-isothermal 
simulations of 4 nsec  with $T=300$ K (i.e., $M_0=1$)
and 100 values of pressure (i.e., $M_1=100$).
Using the obtained histogram $N_{m_0,m_1}(E,\cV)$ of potential 
energy and volume
distribution, we obtained the ST parameters $f_{m_0,m_1}$ by solving the
WHAM equations in Eqs.~(\ref{eqn20p}) and (\ref{eqn21p}). 
We then performed the PST production of 500 nsec and repeated it 10
times with different seeds for random numbers (so, the total simulation
time for the production run is 5.0 $\mu$sec).

%In Fig.~\ref{F5} we show the time series of pressure and potential energy
In Fig.~5 we show the time series of pressure and potential energy
during the PST production run.

%\begin{figure}[th]
\begin{figure}[htbp]
\begin{center}
\includegraphics[width=7.5cm,keepaspectratio]{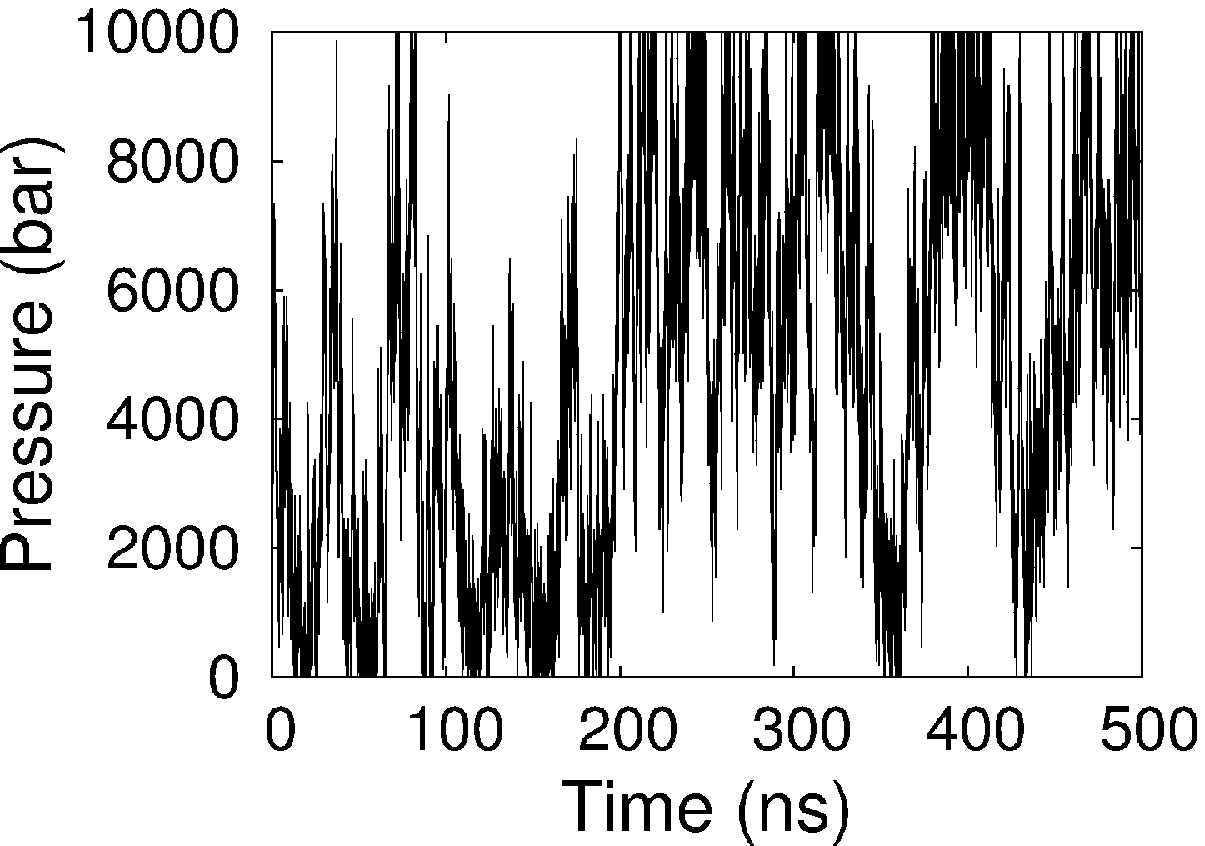}
\includegraphics[width=7.5cm,keepaspectratio]{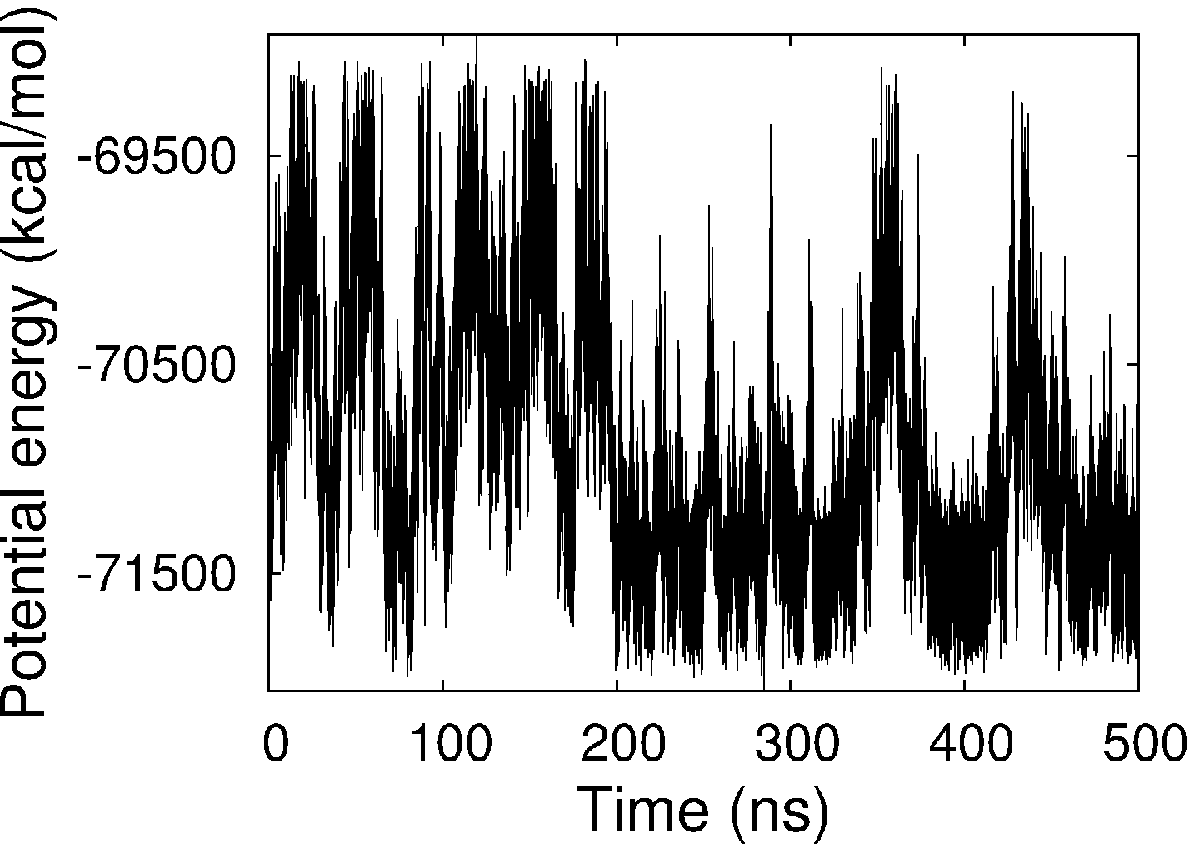}
\caption{Time series of pressure (left) and potential energy (right) during the
PST production run.}
\end{center}
\label{F5}
\end{figure}
In the Figure we see a random walk in pressure between 1 bar and
10000 bar.
A random walk in potential energy is also observed and it 
is anti-correlated with that of pressure, as it should be.

We calculated the fluctuations $\sqrt{<d^2>-<d>^2}$
of the distance $d$ between pairs of $C^{\alpha}$ atoms.
%The results are shown in Fig.~\ref{F6}.
The results are shown in Fig.~6.

%\begin{figure}[th]
\begin{figure}[htbp]
\begin{center}
\includegraphics[width=9.0cm]{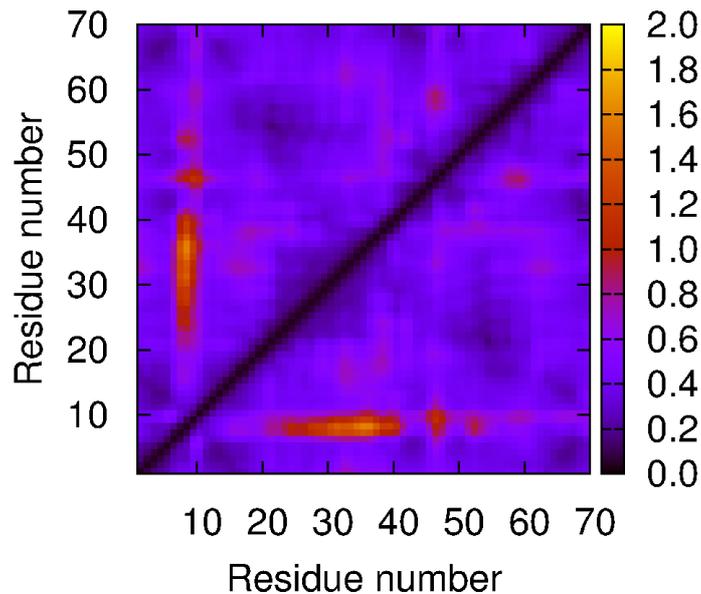}
\caption{Fluctuations of distance between pairs of $C^{\alpha}$ atoms
that was calculated from the PST production run.}
\end{center}
\label{F6}
\end{figure}
We see that large fluctuations are observed between residues
around 7-10 and around 20-40, which are in accord with
the experimental results \cite{KA03,KYA05}.

The flucutating distance corresponds to that between
the turn region of the $\beta$-hairpin and the end of
%$\alpha$-helix as depicted in Fig.~\ref{F7}.
the $\alpha$-helix as depicted in Fig.~7.
While at low pressure this distance is small, at high pressure
it is larger and water comes into the created open region.

%\begin{figure}[th]
\begin{figure}[htbp]
\begin{center}
\includegraphics[width=7.5cm,keepaspectratio]{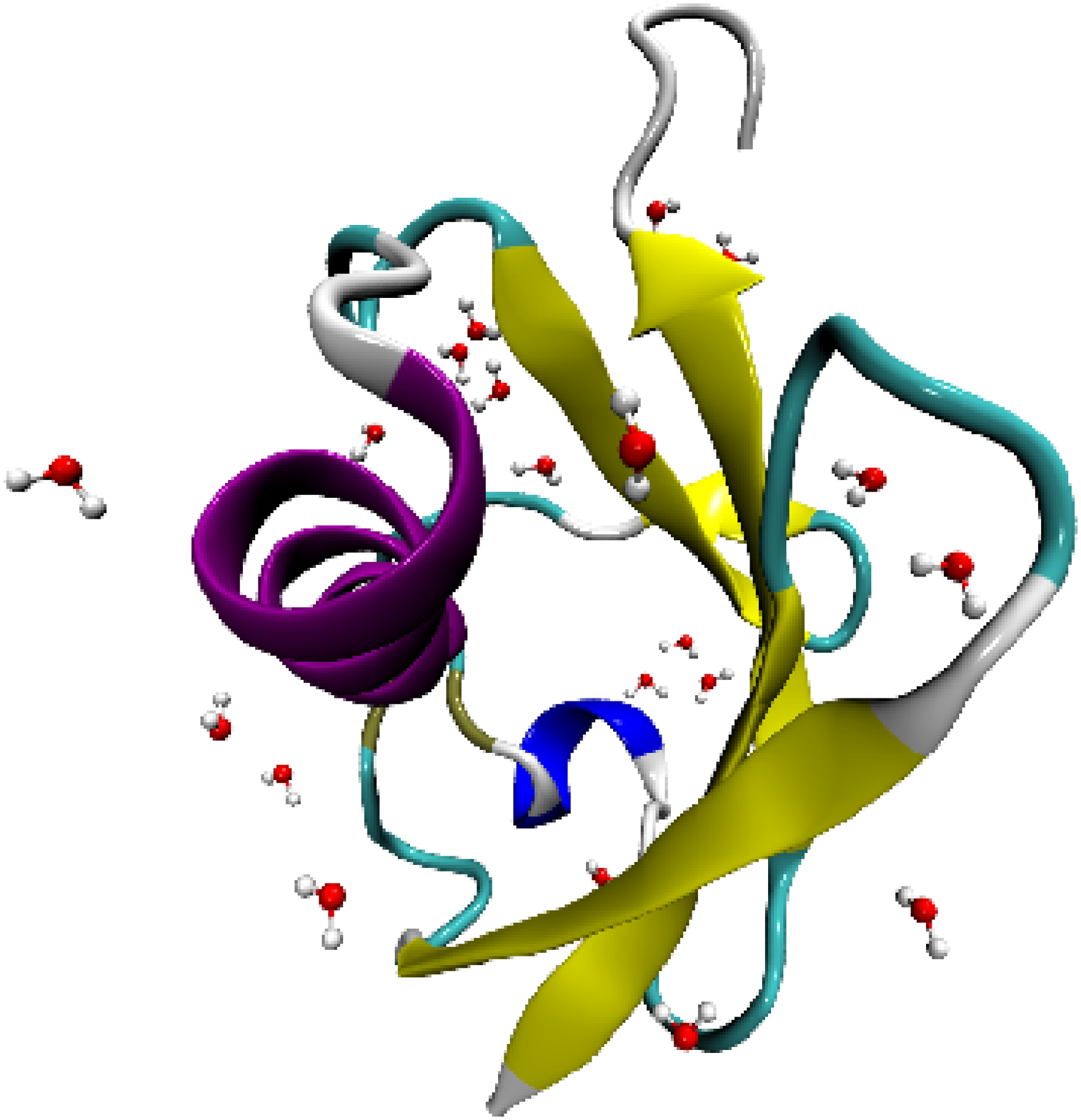}
\includegraphics[width=7.5cm,keepaspectratio]{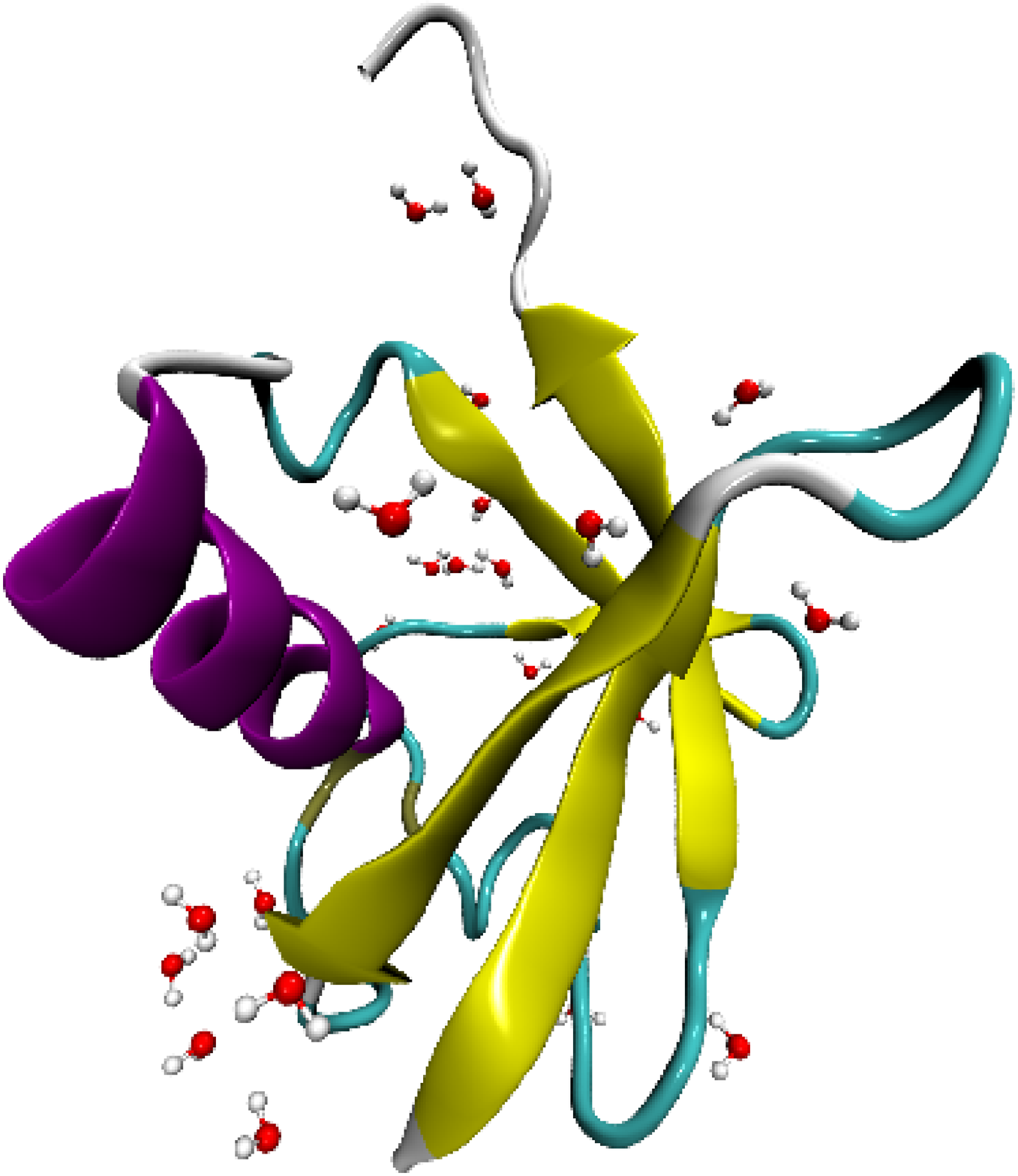}
\caption{Snapshots of ubiquitin during the PST production run
at low pressure (left) and at high pressure (right).}
\end{center}
\label{F7}
\end{figure}

\section{CONCLUSIONS}
In this article we first introduced three well-known generalized-ensemble
algorithms, namely, REM, ST, and MUCA, which can greatly enhance
conformational sampling of biomolecular systems.  We then
presented various extensions of these algorithms.  Examples are
the general formulations of the multidimensional 
REM, ST, and MUCA. 
We generalized the original potential energy function $E_0$ by
adding any physical quantities $V_{\ell}$ of interest as a new
energy term with a coupling constant $\lambda^{(\ell)}$($\ell=1,\cdots,L$). 
The simulations in multidimensional REM
and multidimensional ST algorithms realize a random walk 
in temperature and $\lambda^{(\ell)} (\ell=1,\cdots,L)$ spaces. 
On the other hand, 
the simulation in multidimensional MUCA algorithms realizes a random
walk in $E_0, V_1, \cdots, V_L$ spaces. 

While the multidimensional REM simulation can be easily performed 
because no weight factor determination is necessary, 
the required number of replicas can be quite large and 
computationally demanding. 
We thus prefer to use the multidimensional ST or MUCA, 
where only a single replica is simulated, instead of REM. 
However, it is very difficult to obtain optimal weight factors 
for the multidimensional ST and MUCA. 
Here, we have proposed a powerful method to determine these weight factors. 
Namely, we first perform a short multidimensional REM simulation and 
use the multiple-histogram reweighting techniques to determine the 
weight factors for multidimensional ST and MUCA simulations. 

The multidimensional generalized-ensemble algorithms that 
were presented in the present article will be very useful 
for Monte Carlo and molecular dynamics simulations of 
complex systems such as spin glass, polymer, and biomolecular systems. 

%\newpage
\noindent
{\bf Acknowledgements}: \\
Some of the results were obtained by the computations on the super computers at 
the Institute for Molecular Science, Okazaki, and
the Institute for Solid State Physics, University of Tokyo, Japan. 
This work was supported, in part, by Grants-in-Aid
for Scientific Research on 
Innovative Areas (``Fluctuations and Biological Functions''),
and for the Next-Generation Super Computing Project, Nanoscience Program
from the Ministry of Education, Culture, Sports, Science and Technology (MEXT), Japan.
%\newpage

% BibTeX users please use
% \bibliographystyle{}
% \bibliography{}
%
% Non-BibTeX users please follow the syntax
% the syntax of "referenc.tex" for your own citations
%\input{referenc2}
%%%%%%%%%%%%%%%%%%%%%%%%%%%%%%%%%%%%%%%%%%%%%%%%%%%%%%%%%%%%%%%%%%%%%%  }
%
% BibTeX users please use
% \bibliographystyle{}
% \bibliography{}
%
% Non-BibTeX users please use

%%%%%%%%%%%%%%%%%%%%%%%%%%%%%%%%%%%%%%%%%%%%%%%%%%%%%%%%%%%%%%%%%%%%%%

%\printindex

\end{document}